\documentclass[aps, prl, reprint,10pt,longbibliography,superscriptaddress]{revtex4-2}

\usepackage{upgreek,mathtools}
\usepackage{color}
\usepackage{graphicx}
\usepackage{nccmath}
\usepackage{bm}
\usepackage{hyperref}
\hypersetup{colorlinks,breaklinks,
            urlcolor=[rgb]{0,0,0.36},
            linkcolor=[rgb]{0,0,0.36},
            citecolor=[rgb]{0,0,0.36}}
\usepackage[mathlines]{lineno}
\usepackage{dcolumn}
\usepackage{mathtools}  
\usepackage{dsfont}
\usepackage{comment}
\usepackage{physics}
\usepackage{xcolor}
\usepackage{epsfig,tabularx,multirow,booktabs}

\newcounter{SN}

\newcommand{\ETH}{Institute for Theoretical Physics, ETH Zurich, 8093 Zurich, Switzerland.}

\newcommand{\Palermo}{Dipartimento di Fisica e Chimica “E. Segrè”, Group of Interdisciplinary Theoretical Physics, Università degli Studi di Palermo, I-90128 Palermo, Italy.}

\newcommand{\Vienna}{Vienna Center for Quantum Science and Technology (VCQ), Atominstitut, TU Wien, Vienna, Austria.}

\newcommand{\Salerno}{Dipartimento di Fisica ``E.~R.~Caianiello", Università degli Studi di Salerno, I-84084 Fisciano, Salerno, Italy.}

\begin{document}
\title{Momentum-resolved two-dimensional spectroscopy as a \\probe of nonlinear quantum field dynamics}

\author{Duilio De Santis}
\thanks{These authors contributed equally to this work.}
\affiliation{\Palermo}
\affiliation{\ETH}

\author{Alex~G\'{o}mez~Salvador}
\thanks{These authors contributed equally to this work.}
\affiliation{\ETH}

\author{Nataliia Bazhan}
\affiliation{\Vienna}

\author{Sebastian Erne}
\affiliation{\Vienna}

\author{Maximilian Prüfer}
\affiliation{\Vienna}

\author{Claudio Guarcello}
\affiliation{\Salerno}

\author{Davide Valenti}
\affiliation{\Palermo}

\author{Jörg Schmiedmayer}
\affiliation{\Vienna}

\author{Eugene Demler}
\affiliation{\ETH}

\date{\today}

\begin{abstract}

Emergent collective excitations constitute a hallmark of interacting quantum many-body systems, yet in solid-state platforms their study has been largely limited by the constraints of linear-response probes and by finite momentum resolution. We propose to overcome these limitations by combining the spatial resolution of ultracold atomic systems with the nonlinear probing capabilities of two-dimensional spectroscopy (2DS). As a concrete illustration, we analyze momentum-resolved 2DS of the quantum sine-Gordon model describing the low energy dynamics of two weakly coupled one-dimensional Bose–Einstein condensates. This approach reveals distinctive many-body signatures, most notably asymmetric cross-peaks reflecting the interplay between isolated ($B_2$ breather) and continuum ($B_1$ pair) modes. The protocol further enables direct characterization of anharmonicity and disorder, establishing momentum-resolved 2DS as both a powerful diagnostic for quantum simulators and a versatile probe of correlated quantum matter.

\end{abstract}

\maketitle

\textit{Introduction.---}
Emergent collective excitations---whose properties differ qualitatively from those of their microscopic constituents---are ubiquitous in quantum many-body systems. Prominent examples include fractional excitations in quantum Hall and one-dimensional systems~\cite{Tsui_1982, Laughlin_1983, Haldane_1981_PRL, Haldane_1981, Giamarchi_2003}. Quantum emulators based on ultracold atoms, superconducting qubits, and trapped ions enable controlled realizations of paradigmatic models, offering a unique opportunity to revisit long-standing many-body questions and to develop refined probes of collective excitations~\cite{Bloch_2008, Cazalilla_2011, Langen2015, Gross2017, Monroe2021, King2022}. By enabling the realization of paradigmatic models with tunable parameters, these systems make it possible to revisit long-standing questions in many-body physics with a new degree of control. At the same time, they motivate the development of refined spectroscopic protocols for accessing collective excitations beyond the limits of standard solid-state approaches.

Traditionally, excitations have been probed using linear-response methods such as X-ray scattering~\cite{guinier1994_xray,Comin2016}, optical spectroscopy~\cite{Jimenez2016}, and transport~\cite{robson2017fundamentals,Pekola_heat_transport_2021}. While invaluable, these techniques are often limited by finite momentum resolution. Recent advances in ultracold atomic platforms have allowed some of these linear response methods to be implemented even with momentum resolution~\cite{Gupta_2003_RF_spectroscopy,Endres2012,Prichard2025}, effectively surpassing the limitations of conventional solid-state experiments.
\begin{figure}
    \centering
    \includegraphics[width=0.95\columnwidth]{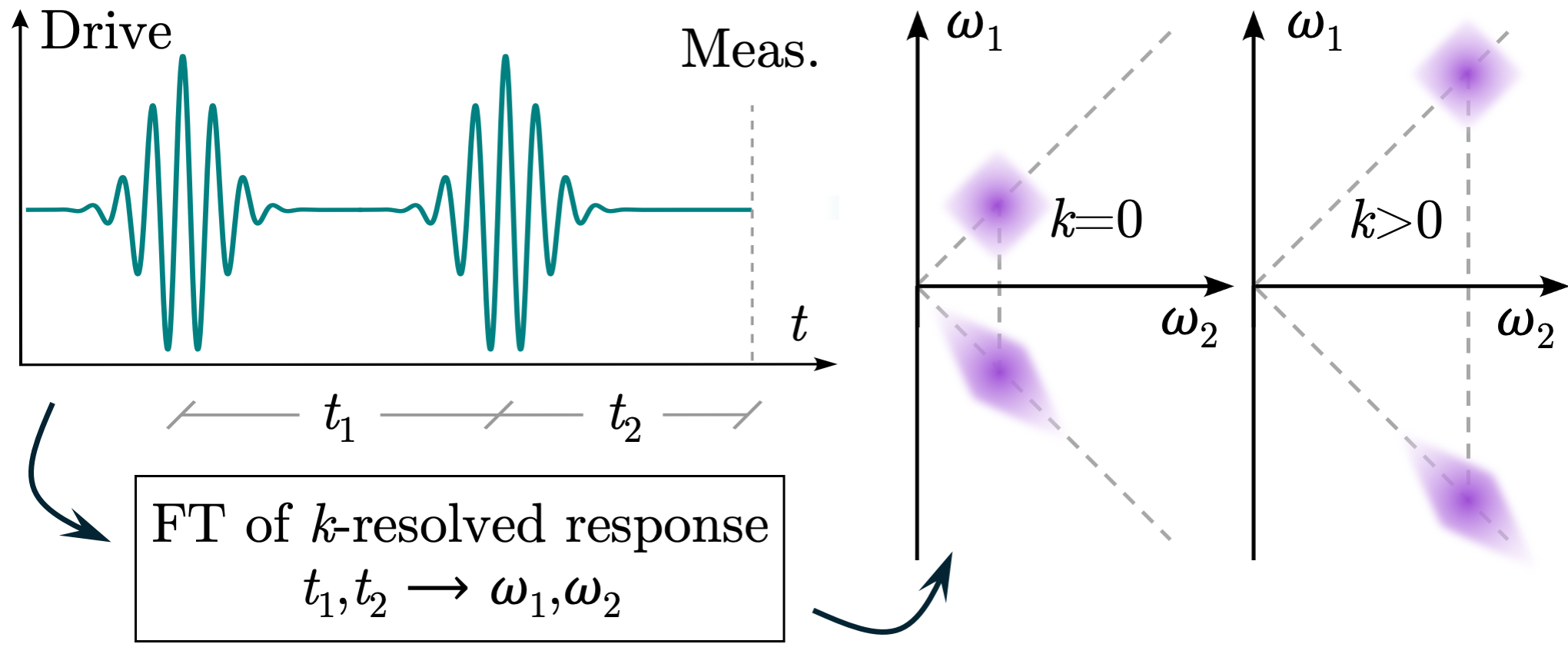}
    \caption{Schematics of a 2DS protocol. Two perturbations, separated by a time delay $t_1>0$, are applied to the system. At a time $t_2>0$ after the second perturbation, the momentum-resolved response is measured. Fourier transforming with respect to the delays $t_1$ and $t_2$ yields a 2D map, for each value of $k$, in frequency domain. The 2D map exhibits several peaks whose positions may depend on $k$, reflecting the momenta of the collective modes under consideration. 
    }
    \label{fig:1}
\end{figure}

Recently, solid-state research has begun to incorporate higher-order spectroscopic techniques~\cite{Autere2018THG_TMD,Seibold2021THG,THG_2024,Liu2025_perspective,huang2025review} to a broad array of systems. In particular, two-dimensional spectroscopy (2DS) has been applied to superconductors~\cite{Mootz2022,Liu_2023_echo,katsumi_revealing_2024_benfatto,puviani2024theory,katsumi2024amplitudemodemultigapsuperconductor,salvador2024principles}, correlated electron systems~\cite{barbalas2023energy,chen2024multidimensionalcoherentspectroscopycorrelated}, ferroics~\cite{Lu2017,Lin2022,zhang2024terahertz1,zhang2024terahertz2,parameswaran2020asymptotically1}, and topological phases~\cite{Blank2023,wan2019resolving,liebman2023multiphotonspectroscopydynamicalaxion,Choi2020SpinLiquid,mcginley2022signatures,McGinley2024Anomalous,potts2024signaturesspinondynamicsphase}, where it enables identification of interaction pathways~\cite{lomsadze2017frequency,liu2018vibrational,muir2022interactions}, classification of bound states~\cite{stone2009two,hao2017trion,hao2017neutral}, and separation of homogeneous from inhomogeneous broadening~\cite{Bristow2011,moody2015intrinsic,huang2023quantum, Liu_2023_echo, salvador2025echo}. Yet, the restricted momentum resolution in traditional solid-state platforms still limits the full power of these approaches. Here, we propose to combine the high momentum resolution of ultracold atoms with the nonlinear capabilities of 2DS, thereby establishing a new paradigm for probing collective excitations, see Fig.~\ref{fig:1} for a schematic of the protocol. A comprehensive discussion on the 2DS technique can be found in the Supplemental Material~\cite{Supplement}.

While our proposal is broadly applicable, we focus here on the sine–Gordon (SG) model, with a twofold motivation encompassing its rich theoretical structure and its experimental relevance.
On the theoretical side, the SG model is an archetype of low-dimensional field theory: it is integrable and hosts soliton collective modes (kinks and breathers) in both its classical and quantum forms~\cite{Dauxois_2006, Torrielli_2024}. It is also dual to the massive Thirring model~\cite{Coleman_1975}, closely connected to Berezinskii–Kosterlitz–Thouless physics~\cite{Giamarchi_2003, Benfatto2007}, and serves as a universal low-energy effective theory for a wide class of low-dimensional systems.
On the experimental side, the SG model is realizable in ultracold atoms, where it captures the low-energy dynamics of two weakly coupled one-dimensional Bose–Einstein condensates~\cite{Schweigler2017, Schweigler2021}. Beyond coupled BECs, the SG framework finds application extended Josephson junctions and high-temperature superconductors~\cite{Jesus_2014, michael2024giant, DeSantis2025}, spin systems, charge-density waves, disordered systems~\cite{Lee_1974, Rice_1976, Asano_2000, Giamarchi_2003, Tiegel_2016, Visuri_2020, Wybo_2022, Wybo_2023}, multicomponent quantum Hall systems~\cite{Girvin_1995}, domain boundaries in anisotropic magnetic systems~\cite{Zharnitsky_1998}, and state-of-the-art quantum processors~\cite{Andersen2024}.

Our objective is to demonstrate that momentum-resolved 2DS provides a powerful probe for quantum simulators. Using the SG model as a representative interacting system, we theoretically implement nonlinear response spectroscopy within a 2DS protocol. The resulting 2D maps reveal characteristic many-body features, including asymmetric cross-peak patterns arising from the coupling of isolated and continuum modes. Additionally, these protocols enable the characterization of interaction pathways, anharmonicity, and shot-to-shot disorder.

\vspace{0.25cm}
\textit{Model and driving protocol.---}
We write, in dimensionless units, the SG Hamiltonian as
\begin{equation}
H_0 = \frac{1}{2} \int dx \left\lbrace \left( \partial_x \varphi \right)^2 + \Pi^2 -\Delta_0  \cos (\beta \varphi) \right\rbrace ,
\label{eqn:sG_H}
\end{equation}
where the fields ${ \varphi }$ and ${ \Pi }$ satisfy bosonic commutation relations $\comm{\varphi(x)}{\Pi(x')}=i\delta(x-x')$, and $\beta$ and $\Delta_0$ are free parameters~\cite{Gritsev2007}. Equation~\eqref{eqn:sG_H} is among the most universal nonlinear theories, and its exact diagonalization unveils a rich spectrum of excitations depending on the value of $\beta$: for $0<\beta^2<4\pi$, both (anti)kinks and their bound states, breathers, are found; for $4\pi<\beta^2<8\pi$, no breather can form; finally, for $\beta^2>8\pi$, the cosine term becomes irrelevant in the renormalization group sense and the Luttinger liquid model is recovered~\cite{Malard_2013}. 
The parameter $\beta$ is connected to the Luttinger constant $K$ by $\beta^2=2\pi/K$, and $\Delta_0 \sim J $ is the bare coupling associated with the tunneling strength $J$ between the two condensates~\cite{Supplement}.

\begin{figure}
    \centering
    \includegraphics[width=0.9\linewidth]{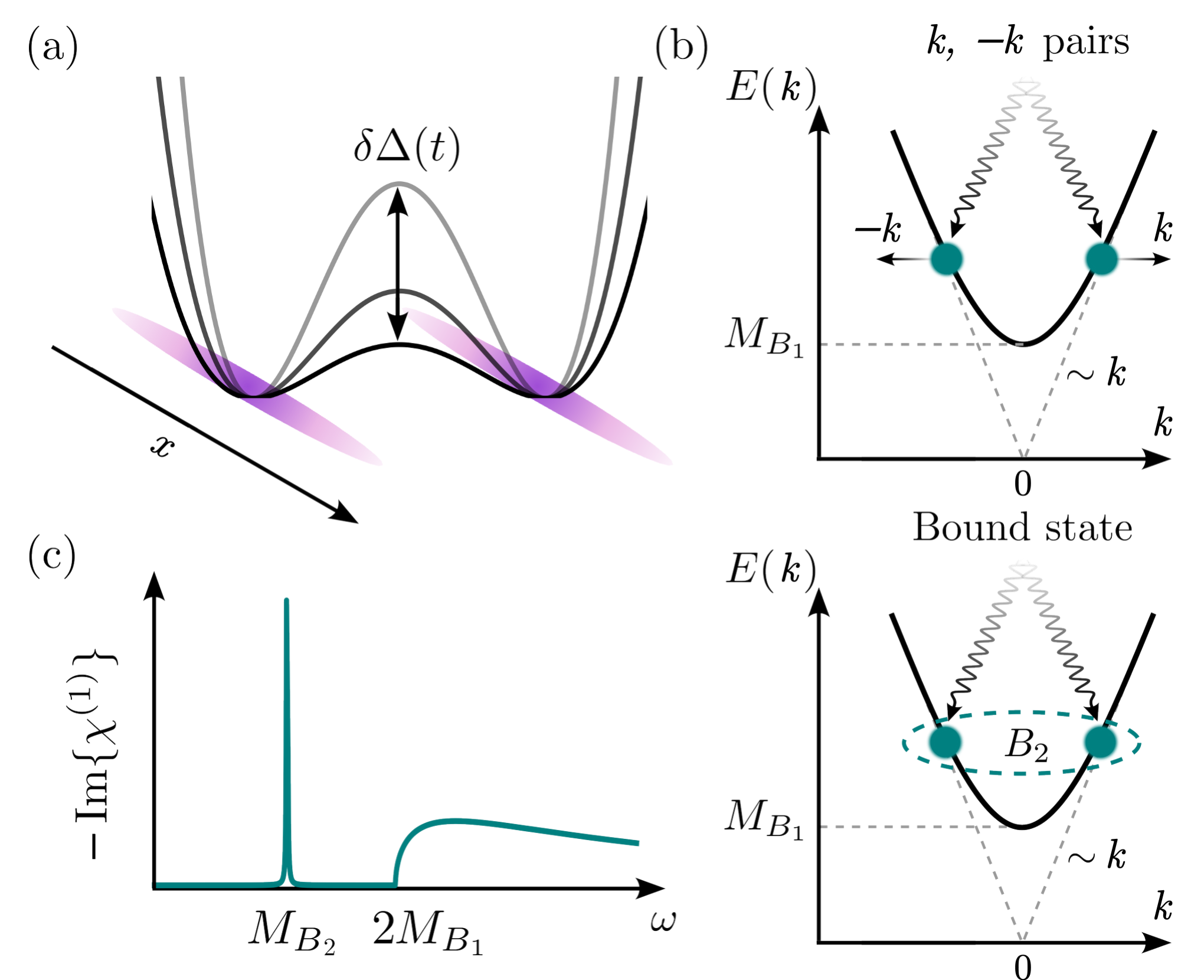}
    \caption{(a) Pictorial representation of the external modulation of the tunneling strength, see Eq.~\eqref{eqn:modulation}, in the system of coupled condensates. (b) Sketch of the relevant excitations accessible via tunneling modulation within our theory. (c) Imaginary part of the linear response function at total momentum $k=0$. From left to right we distinguish two types of excitations: a sharp resonance corresponding to the $B_1-B_1$ bound state, labelled as $B_2$, and a two-particle continuum composed of the $(k, -k)$ $B_1$ pairs.}
    \label{fig:2}
\end{figure}

Our aim is to investigate the response of the SG system to a spatially homogeneous modulation of the coupling $\Delta_0$ given by~\footnote{A generalization to spatially dependent scenarios~\cite{Gritsev2007} is straightforward. We also note that while our subsequent results are derived for a generic $\delta\Delta_{0}(t)$, for plotting purposes we mostly focus on impulsive (delta-like) forcing profiles.}
\begin{equation}
    H_{d} = -\frac{1}{2} \int dx \,\delta\Delta_{0} (t)  \cos (\beta \varphi),
\label{eqn:modulation}
\end{equation}
%
which corresponds, within the coupled BECs realization, to a time-dependent modulation of the tunneling strength [see Fig.~\ref{fig:2}(a)].
Note that, since both $H_0$ and $H_d$ are even in the field $\varphi$, only even-power correlators can acquire expectation value. 

Henceforth, we assume a self-consistent Gaussian Ansatz~\cite{SHI2018245}, which captures effects beyond the harmonic approximation, to characterize the dynamics of the system. This approach replaces the interacting cosine potential with an effective quadratic term whose mass is dynamically determined in a self-consistent way. In the low-$T$, low-$\beta$ regime, the framework reliably tracks the lowest-lying SG breather excitations~\cite{Supplement}, which usually dominate the system's response. More precisely, we have access to two distinct mode contributions, as depicted in Fig.~\ref{fig:2}(b) and (c). The first corresponds to pairs of $B_1$ breathers with opposite momenta $k$ and $-k$, which appear as a continuum in the spectral function, $-\text{Im}\lbrace \chi^{(1)} \rbrace$, above the two-particle excitation gap $2M_{B_1}$. The second corresponds to the $B_2$ breather at zero momentum, manifesting as an isolated peak in $-\text{Im}\lbrace \chi^{(1)} \rbrace$, and it constitutes the lowest energy excitation accessible within our driving protocol~\cite{Gritsev2007}. 

\begin{figure*}
    \centering
    \includegraphics[width=\textwidth]{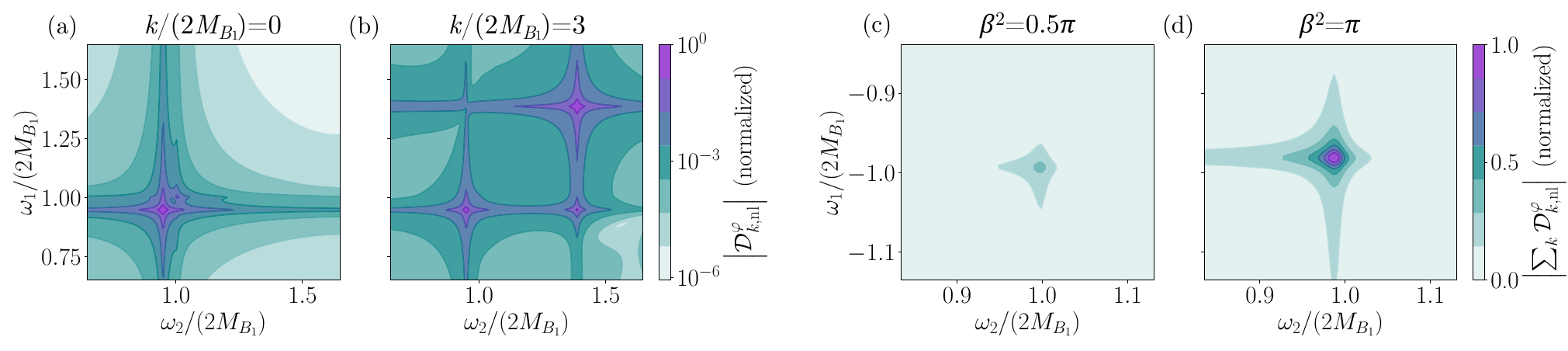}
    \caption{(a)-(b) The (normalized) $k$-resolved absolute value 2D map, $ | \mathcal{D}_{k,\rm{nl}}^{\varphi} | $. Taking $\beta^2 = 2 \pi$, we explore the system's nonrephasing sector for zero $k$ [panel (a)] and for $ k / ( 2 M_{B_1} ) = 3 $ [panel (b)]. Panels (a) and (b) share the same logarithmic color scale. As $k$ increases, the interplay between the $B_1$ continuum and the isolated $B_2$ becomes apparent, leading to asymmetric cross-peaks. This asymmetry is a hallmark of many-body dynamics absent in standard coupled-oscillator models. (c)-(d) The (normalized) absolute value summed over $k$ 2D map, $ | \sum_k \mathcal{D}_{k,\rm{nl}}^{\varphi} | $. We focus on the system's rephasing sector for $\beta^2 = 0.5 \pi$ [panel (c)] and $\beta^2 = \pi$ [panel (d)]. In both panels (c) and (d), the intensity is normalized to the $\beta^2 = \pi$ maximum (the colorbar is shared). As $\beta^2$ increases, the system progressively becomes more interacting and the rephasing peak more prominent. In panels (a)-(d), frequencies are in units of the $2 B_1$ mass, and a value $\eta = 0^+$ is chosen for the damping parameter. }
    \label{fig:3}
\end{figure*}

Formally, our analysis focuses on the two-point correlators, which we define in terms of the Fourier-transformed fields $\varphi_k$ and $\Pi_k$:
\begin{align}
    {\cal D}_{k}^{\varphi} (t) &=\langle \varphi_{-k}(t) \varphi_k(t) \rangle , \notag \\ 
    {\cal D}_{k}^{\Pi} (t) &=\langle\Pi_{-k}(t) \Pi_k(t) \rangle , \label{eqn:correlators} \\
    {\cal D}_{k}^{X} (t) &=\frac{\expval{\acomm{\Pi_{-k}(t)}{\varphi_k(t)}} + \expval{\acomm{\varphi_{-k}(t)}{\Pi_k(t)}}}{2} . \notag  
    \label{eqn:correlators}
\end{align}
Note that since the SG Hamiltonian is translationally invariant, and the external drive is spatially homogeneous, Eq.~\eqref{eqn:correlators} covers the entire set of nonzero two-point correlators. In the following, we analyze the second-order response of the system to the external perturbation described in Eq.\eqref{eqn:modulation}, using a scheme similar to that of Ref.~\cite{salvador2024principles}. Specifically, a perturbative expansion of $\mathcal{D}^{\alpha}_k$ ($\alpha=\varphi,\Pi,X$) in powers of $\delta\Delta_0$ is performed up to second order 
\begin{equation}
    \mathcal{D}^{\alpha}_k (t) = \mathcal{D}^{\alpha}_{k,0} + \mathcal{D}^{\alpha}_{k,1}(t) + \mathcal{D}^{\alpha}_{k,2}(t) + \dots ,
\end{equation}
where the sub-index $i$ in $\mathcal{D}^{\alpha}_{k,i}$ represents the order in $\delta\Delta_0$, and $i=0$ corresponds to the equilibrium thermal fluctuations. For completeness, a brief discussion of the system's linear response can be found in~\cite{Supplement}. Before proceeding, we point out that, as expected from a low-energy quantum field theory, the equilibrium fluctuations computed within the Gaussian Ansatz diverge. Given that the fundamental bosons of the SG theory correspond to $B_1$ breathers~\cite{gritsev2007spectroscopy}, a regularization is performed by defining the physical mass $M_{B_1}$ and drive $\delta\Delta$ as opposed to the bare couplings $\Delta_0$ and $\delta\Delta_0$. Details on our regularization and computation of $\mathcal{D}_{k,0}^\alpha$ can be found in~\cite{Supplement}.

\vspace{0.25cm}
\textit{2D spectroscopy of fluctuation dynamics.---}
In the 2DS framework, two perturbations $\delta\Delta_{\rm A}$ and $\delta\Delta_{\rm B}$ separated by a time delay $t_1>0$ are applied onto the system, see Fig.~\ref{fig:1}. At a time $t_2>0$ after the second perturbation, the momentum resolved variance $\mathcal{D}_{k,\rm AB}^{\varphi}$ is measured. A purely nonlinear response $\mathcal{D}_{k,\rm nl}^{\varphi}$ can be obtained by subtracting from $\mathcal{D}_{k,\rm AB}^{\varphi}$ the contributions arising from single perturbations, such that
\begin{equation}
    \mathcal{D}_{k,\rm nl}^\varphi(t_1,t_2) = \mathcal{D}_{k,\rm AB}^\varphi(t_1,t_2) - \mathcal{D}_{k,\rm A}^\varphi(t_1,t_2) - \mathcal{D}_{k,\rm B}^\varphi(t_1,t_2),
\end{equation}
where $\mathcal{D}_{k,\rm A}^\varphi$ and $\mathcal{D}_{k,\rm B}^\varphi$ are the measured variances when only perturbation $\delta\Delta_{\rm A}$ and $\delta\Delta_{\rm B}$ are present, respectively. Note that, for this system and protocol, we are probing the second order response $\chi_k^{(2)}$, and that the nonlinear signal is directly proportional to $\delta\Delta_{\rm A}\delta\Delta_{\rm B}$. The nonlinear response, $\mathcal{D}_{k,\rm nl}^\varphi(t_1,t_2)$, is subsequently Fourier-transformed with respect to $t_1$ and $t_2$, under the constraint $t_1,t_2>0$. This results in $\mathcal{D}_{k,\rm nl}^\varphi(\omega_1,\omega_2)$---what we refer to as the 2D map---which is directly related to $\chi^{(2)}$ via 
\begin{align}
    \mathcal{D}_{k,\rm nl}^\varphi(t_1,t_2)=2\int\frac{d\omega_1}{2\pi}
    \int\frac{d\omega_2}{2\pi}\chi^{(2)}_k(\omega_1+\omega_2,\omega_1)\notag \\ \times  e^{-i\omega_1(t_1+t_2)}e^{-i\omega_2 t_2}\delta\Delta_{\rm A}(\omega_1)\delta\Delta_{\rm B}(\omega_2).
\end{align}
%
We provide details of the calculation of $\chi^{(2)}$ within our Gaussian state Ansatz, or by means of an equivalent diagrammatic approach, in~\cite{Supplement}.

Our protocol deviates from more common 2D approaches~\cite{Liu_2023_echo,salvador2024principles,salvador2025echo} in two main ways. First, the external perturbation couples to operators with even powers of the bosonic field, i.e., $\varphi^{2n}$, rather than to $\varphi$. This change has an important consequence: it produces a finite second-order response in powers of $\delta\Delta$, even when the system itself has no intrinsic nonlinearities. Second, the perturbation does not act on a single operator but instead couples to an infinite series of operators $\propto \cos(\beta\varphi)$, as shown in Eq.~\eqref{eqn:modulation}. This richer structure generates distinctive $\beta$-dependent features in the 2D map, which we illustrate in the following. 

Before we proceed, it is helpful to note that, for small $\beta$, the pump-probe sector ($\omega_1=0$) of the resulting maps is featureless, i.e., it shows no peaks. We have verified this by deriving the two-sided Feynman diagram rules for our protocol with a single bosonic excitation, see Supplemental Material~\cite{Supplement}. Therefore, the discussion of the 2D spectra will focus on the nonrephasing and rephasing sectors, which correspond to the $\omega_1\omega_2>0$ and $\omega_1\omega_2<0$ quadrants of the 2D maps, respectively~\cite{hamm_zanni_2011}. Note that the $\omega_1\omega_2$ sign indicates whether the two consecutive perturbations excite the system in-phase (nonrephasing) or out-of-phase (rephasing).

We begin by considering the signatures in the the nonrephasing sector of the 2D map ($\omega_1\omega_2>0$) as a function of $\beta$. 
Figure~\ref{fig:3}(a)-(b) present the 2D maps as a function of $k$ and $\beta^2 = 2 \pi$. For $k = 0$, see Fig.~\ref{fig:3}(a), our choice of $\beta$ is large enough for the $B_2$ binding energy to be appreciable, resulting in two diagonal peaks: at the $2 B_1$ frequency for the $B_1$ pair, and below the two-particle gap for the $B_2$ bound state. As $k$ increases, see Fig.~\ref{fig:3}(b), the overall peak spacings increase, and off-diagonal features become apparent. 
At first glance, our situation could seem analogous to the study of coupled molecules, where the 2D map provides information about their coupling strength, excitation pathways, and even Fermi resonances~\cite{hamm_zanni_2011}. One could thus expect the characteristic four peak structure of coupled modes to manifest in our spectra. However, it is worth noting that the $B_1$ pairs constitute a continuum while the $B_2$ is a sharp, isolated excitation. The presence of this continuum leads to the suppression of one of the cross-peaks due to dephasing, resulting in the qualitatively distinct signature shown in Fig.~\ref{fig:3}(b), see~\cite{Supplement} for more details and schematics. Importantly, this feature is unique to the many-body nature of the problem, as it directly reflects the interplay between a continuum of excitations and an isolated mode. Similar peak patterns have recently been reported in the context of magnetic systems~\cite{Watanabe2025}.

\begin{figure}
    \centering
    \includegraphics[width=\columnwidth]{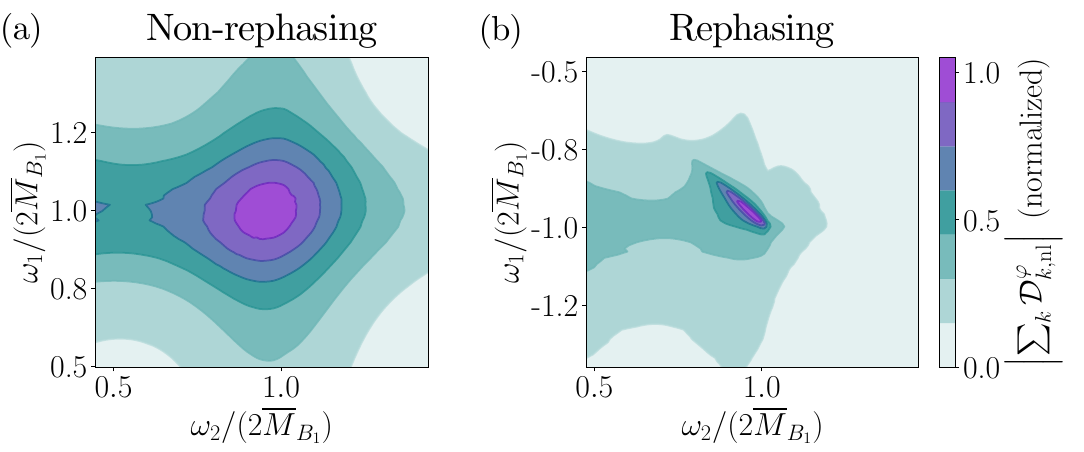}
    \caption{The (normalized) summed over $k$ absolute value 2D map, $ | \sum_k \mathcal{D}_{k,\rm{nl}}^{\varphi} | $. Taking $\beta^2 = \pi$, we average over a shot-to-shot distribution assumed to be Gaussian for $2 M_{B_1}$, with mean $2 \overline{M}_{B_1} $ and standard deviation $\sigma / (2 \overline{M}_{B_1}) = 0.1 $. We focus on the system's non-rephasing [rephasing] sector in panel (a) [panel (b)]. Frequencies are normalized to the average $2 B_1$ mass, a value $\eta = 0^+$ is chosen for the damping parameter, and the colorbar is shared. On the one hand, the non-rephasing peak, see panel (a), becomes symmetrically broadened due to presence of shot-to-shot noise, with a shape indistinguishable from that of a damped mode. On the other hand, the almond-shaped rephasing peak in panel (b) provides a direct diagnostic of shot-to-shot density fluctuations, distinguishing them from intrinsic damping. 
    }
    \label{fig:4}
\end{figure}

We now turn to the rephasing sector ($\omega_1\omega_2<0$), whose name reflects its connection to the spin-echo phenomenon~\cite{Hahn_1950, Bruun_polaron_Echo_2015}. For $\beta \to 0$, it is devoid of any signatures---a distinctive feature of our protocol. Such behavior can be intuitively understood by approximating the external perturbation as $\cos(\beta \varphi)\sim\varphi^2$, and noting that no pathway can lead to echo-type processes at order $\left(\delta\Delta\right)^2$, see Supplemental Material for a detailed discussion~\cite{Supplement}. This is, however, not the case for larger $\beta$, where the where the cosine potential has to be considered. As such, the presence of the echo peak in the 2D spectra is a direct sign of a non-harmonicity of the system---equivalently, an indicator of the strength of interactions of the theory. We present the evolution of the echo peak as a function of $\beta$ in Fig.~\ref{fig:3}(c)-(d), highlighting its progressive appearance as $\beta$ increases~\footnote{We note that Fig.~\ref{fig:3}(c)-(d) presents the system's response summed over all $k$, where the bound state is expected to dominate (as indicated by the linear response). This accounts for the appearance of only a sub-$2 B_1$ peak in this case. Rephasing phenomena are observed, in a similar fashion, in the $k$-resolved case as well.}. 

Furthermore, the echo peak is renowned for its capability to disentangle different sources of broadening. The paradigmatic example is in isolated two-level systems, where the echo peak is capable of disentangling and quantifying the homogeneous and inhomogeneous broadening~\cite{Siemens_2010}. Similar effects have been observed in exciton systems~\cite{Bristow2011,moody2015intrinsic,huang2023quantum} and, more recently, in strongly correlated superconducting cuprates~\cite{Liu_2023_echo}.

Of particular interest for the considered ultracold-atom system is the boson density, which determines $\beta$ and $\Delta$, thereby influencing the breather masses and the peak positions in the 2D map. Since observables in ultracold-atom experiments are typically obtained by averaging multiple destructive measurements, shot-to-shot fluctuations in the boson density---uncorrelated between distinct realizations---are expected. These fluctuations can affect the realization-averaged echo peak linewidths, in analogy to isolated two-level systems~\cite{Siemens_2010}. We investigate this aspect in Fig.~\ref{fig:4}, where we present the non-rephasing and rephasing (echo) peaks assuming a Gaussian distributed shot-to-shot density with mean $2 \overline{M}_{B_1} $ and standard deviation $\sigma / (2 \overline{M}_{B_1}) = 0.1 $. The non-rephasing peak, see Fig.~\ref{fig:4}(a), symmetrically broadens with a shape indistinguishable from that of a damped mode. The echo peak instead presents a characteristic almond signature, see Fig.~\ref{fig:4}(b). Its extent along the $-\omega_1=\omega_2$ diagonal roughly determines the shot-to-shot variance, while its broadening along $\omega_1=\omega_2$ quantifies the intrinsic lifetime of the excitation. The distinct profiles of the non-rephasing and echo peaks motivate adopting NMR-inspired phenomenological fitting forms~\cite{Siemens_2010} to accurately determine damping and disorder in the ultracold atomic setup.


\vspace{0.25cm}
\textit{Conclusions and outlook.---}
We investigated the nonlinear response of the SG model through a novel 2D spectroscopy protocol. Our approach can be used to access the momentum-resolved nonlinear response functions, leveraging the spatial resolution of present-day ultracold atom platforms.
The nonrephasing sector captures the interplay between continuum and discrete excitations, yielding asymmetric cross-peaks unique to the SG many-body dynamics. In contrast, the rephasing (echo) sector serves as a sensitive probe of interaction strength and of disorder, disentangling damping from shot-to-shot fluctuations.

The signatures predicted here can be directly tested with present atom-chip experiments on tunnel-coupled 1D Bose gases. Barrier modulation provides the required drive, while matter-wave interference gives access to momentum-resolved phase correlations. Typical system sizes, imaging resolution, and tunneling rates ensure that the relevant timescales lie well within condensate coherence times. Thus, the asymmetric cross-peaks and echo-like disorder diagnostics we identify are experimentally accessible with current technology. Although a detailed implementation lies beyond the scope of this paper, some practical considerations are outlined in~\cite{Supplement}.

We note that our nonlinear spectroscopic analysis---presented at zero-temperature---can be seamlessly adapted to incorporate finite-temperature effects by employing the self-consistent mean-field approximation. Furthermore, owing to the integrability of the SG model~\cite{Gritsev2007}, our 2D spectroscopy approach can be extended to account for the full nonlinear effects of the theory. To this end, the nonrephasing results from our protocol strongly indicate its suitability for studying matrix elements between breathers, building on the well-known form factors of the integrable theory $\sim\bra{B_n}e^{i\beta\varphi}\ket{0}$~\cite{Gritsev2007}. In this regard, we observe that the off-diagonal peak in Fig.~\ref{fig:3}(b) is proportional to well-known from factors and $ \bra{B_2(0)}\cos{\beta\varphi}\ket{B_1(k)B_1(-k)}$, while the diagonal ones correspond to $ \bra{B_2(0)}\cos{\beta\varphi}\ket{B_2(0)}$ and $ \bra{B_1(k)B_1(-k)}\cos{\beta\varphi}\ket{B_1(k)B_1(-k)}$~\footnote{Note that in Fig.~\ref{fig:3} we plot the absolute-value peak, but both real and imaginary parts are in principle experimentally accessible, enabling their full characterization.}.

The combination of ultracold atomic systems and multidimensional spectroscopy opens the door to envisioning additional, more sophisticated protocols. For example, a scheme where the perturbation couples linearly to the field $\varphi$ can go beyond the usual $k = 0$ coupling found in condensed matter platforms, enabling momentum-dependent coupling and resolution, and granting full access to the theory's $T$-matrix.
Beyond the SG model, the same framework can be applied to other effective field theories realized in ultracold atoms (Lieb–Liniger gases, spin chains) and to engineered quantum devices (superconducting resonators, trapped ions). By accessing nonlinear matrix elements and form factors, momentum-resolved 2DS provides a route to directly test integrability, explore thermalization, and characterize exotic excitations in quantum simulators.

We thank D. Abanin, M. Aidelsburger, T. Andersen, I. Bloch, A. Carollo, V. Gritsev, F. Marijanovic, M. Michael, I. Morera, L. Skolc, B. Spagnolo, and A. Tsvelik for insightful discussions. A.G.S. and E.D. acknowledge the Swiss National Science Foundation (Grant Number 200021\textunderscore212899) and ETH-C-06 21-2 equilibrium Grant with project number 1-008831-001 for funding. E.D. also acknowledges ARO grant W911NF-20-1-0163 and support from the CORI Project 2024 for the launch and development of international collaborations at the University of Palermo, which supported a scientific visit during which part of this work was carried out. D.D.S. acknowledges support from the Italian Ministry of University and Research (MUR), the “Angelo Della Riccia” Foundation, and ETH Zurich. N.B. and S.E. acknowledge support by the Austrian Science Fund (FWF) [Grant No. I6276, QuFT-Lab]. M.P. has received funding from Austrian Science Fund (FWF) [10.55776/ESP396; QuOntM].



%


\appendix
\clearpage
\newpage
\setcounter{figure}{0}
\setcounter{equation}{0}
\setcounter{page}{1}

\renewcommand{\thepage}{S\arabic{page}} 
\renewcommand{\thesection}{S\arabic{section}} 
\renewcommand{\thetable}{S\arabic{table}}  
\renewcommand{\thefigure}{S\arabic{figure}} 
\renewcommand{\theequation}{S\arabic{equation}} 

\onecolumngrid

\begin{center}
\textbf{\Large{\large{Supplemental Material \\ }}}
\end{center}


\section{Coupled 1D condensates as a sine-Gordon experimental realization}
\label{sec:model}
A prototypal experimental realization of the quantum sine-Gordon (SG) model, first proposed in Ref.~\cite{Gritsev2007}, is that two tunnel-coupled (quasi) 1D condensates. Following Refs.~\cite{Schweigler2017,Schweigler2021}, we consider a system of ultracold bosons (Rubidium-87 atoms) trapped in a double-well potential on an atom chip. The double-well potential, engineered via rf magnetic fields, features a barrier which can be tuned by adjusting such fields' amplitude. In other words, the tunneling rate between the condensates is a highly controllable quantity, a fact which motivates our choice for the driving protocol made in this work. Then, matter-wave interference grants access to the spatially-resolved phase difference between the superfluids. This phase difference can be Fourier-transformed into momentum space and averaged over multiple experimental runs. The resulting data under a single perturbation (or two consecutive time-delayed perturbations) provide direct experimental counterparts to the 1D spectra (or 2D maps) discussed in the main text. See below for further details.

In typical experimental setups~\cite{Schweigler2017,Schweigler2021}, both wells are confined tightly (weakly) in the radial (longitudinal) direction, such that the system is effectively one-dimensional. In this regime, the radial dynamics is frozen, while the double-well barrier still allows for tunneling between the two condensates. After integrating out the radial degrees of freedom, one obtains the following 1D effective Hamiltonian
\begin{equation}
H = \sum_{j=1}^2 \int dx \left[ \frac{\hbar^2}{2 m} \partial_x \psi_j^\dagger \partial_x \psi_j + \frac{g_{\rm{1D}}}{2} \psi_j^\dagger \psi_j^\dagger \psi_j \psi_j + U(x) \psi_j^\dagger \psi_j - \mu \psi_j^\dagger \psi_j \right] - \hbar J \int dx \left[ \psi_1^\dagger \psi_2 + \psi_2^\dagger \psi_1 \right] ,
\end{equation}
where the field operators fulfill bosonic commutation relations, ${ \comm{\psi_j (x)}{\psi_{j'}^\dagger (x')}= \delta_{jj'} \delta(x-x') }$, $m$ is the atomic mass, $g_{\rm{1D}}$ the effective 1D interaction strength, $U(x)$ the longitudinal trapping potential (set to zero in the following for simplicity), and $J$ the tunneling rate.

The effective coupling $g_{\rm{1D}}$ can be expressed in terms of the $s$-wave scattering length $a_s$ and the 1D density $n_{\rm{1D}}$ as
\begin{equation}
g_{\rm{1D}} = \hbar \omega_\perp a_s \frac{2 + 3 a_s n_{\rm{1D}}}{1 + 2 a_s n_{\rm{1D}}} \, .
\end{equation}
Working in the density–phase representation,
\begin{equation}
\psi_j (x) = e^{i \theta_j (x)} \sqrt{ n_{\rm{1D}} + \delta n_j (x) } ,
\end{equation}
we introduce the relative degrees of freedom
\begin{equation}
\varphi (x) = \theta_1 (x) - \theta_2 (x) , \qquad \rho (x) = \frac{1}{2} \left[ \delta n_1 (x) - \delta n_2 (x) \right],
\end{equation}
which satisfy canonical commutation relations. Expanding the Hamiltonian to second order in $ \partial_x \varphi $ and $ \rho $, one finds
\begin{equation}
H = \int dx \left[ \frac{\hbar^2 n_{\rm{1D}}}{4 m} \left( \partial_x \varphi \right)^2 + \frac{\hbar^2}{4 m n_{\rm{1D}}} \left( \partial_x \rho \right)^2  + g \rho^2 - 2 \hbar J n_{\rm{1D}} \cos \varphi \right] ,
\end{equation}
with $ g = g_{\rm{1D}} + \hbar J / n_{\rm{1D}} $. Since density fluctuations are strongly suppressed in the low-energy regime of interest, the density gradient term may be neglected, yielding the sine-Gordon (SG) Hamiltonian
\begin{equation}
H = \int dx \left[ \frac{\hbar^2 n_{\rm{1D}}}{4 m} \left( \partial_x \varphi \right)^2 + g \rho^2 - 2 \hbar J n_{\rm{1D}} \cos \varphi \right] .
\end{equation}
For $J=0$ this reduces to the Luttinger Hamiltonian. Introducing the rescaled fields ${ \varphi \to \varphi / \beta }$ and ${ \Pi = \beta \rho }$, with ${ \beta^2 = 2 \pi / K }$ and $K$ the Luttinger parameter, the Hamiltonian becomes
\begin{equation}
H = \frac{\hbar c_s}{2} \int dx \left[ \left( \partial_x \varphi \right)^2 + \Pi^2 - \frac{4 J n_{\rm{1D}}}{c_s} \cos \left( \beta \varphi \right) \right] ,
\end{equation}
where ${ c_s = \sqrt{g n_{\rm{1D}} / m } }$ is the sound velocity. Finally, introducing a reference length scale $x_0$, one can write
\begin{equation}
H = \frac{\hbar c_s}{2 x_0} \int d x \left[ \left( \partial_x \varphi \right)^2 + \Pi^2 - \Delta \cos \left( \beta \varphi  \right) \right] ,
\end{equation}
with the dimensionless parameter ${ \Delta = 8 J m x_0^2 / \left( \beta^2 \hbar \right) }$.

\textit{Typical experimental parameters.---} In cold-atom realizations with $^{87}$Rb~\cite{Schweigler2017,Schweigler2021}, typical values are: radial trapping frequencies of order ${\omega_\perp / 2\pi \sim \text{kHz}}$ and longitudinal frequencies of order ${\omega_x / 2\pi \sim \text{few Hz}}$; interaction energies ${\mu/\hbar}$ and thermal energies ${k_B T/\hbar}$ in the few hundred Hz range, both below $\omega_\perp$, ensuring effective one-dimensionality; $s$-wave scattering length $a_s \approx 5\,\text{nm}$; mean 1D densities $n_{\rm 1D} \sim 10$--$100 \,\mu\text{m}^{-1}$; and tunneling rates $J$ ranging from a few to a few tens of Hz. For a system size of order $L \sim 50\,\mu\text{m}$ and an imaging resolution of a few micrometers, the accessible momentum range is approximately $0.1$ to $3\,\mu\text{m}^{-1}$. These values illustrate that the assumptions made in the derivation of the effective SG description are well justified in typical experiments.

\textit{Practical implementation and measurements.---} In tunnel-coupled 1D Bose gases on atom chips (see~\cite{Schweigler2017,Schweigler2021}), the modulation $\delta\Delta(t)$ used in our theory corresponds to a controlled change of the barrier height. In practice, this is achieved by tuning the amplitude of the radio-frequency or optical fields that generate the double-well potential. Such modulations on millisecond timescales are standard and have already been demonstrated experimentally.

The relevant observables are directly accessible: matter-wave interference yields the spatially resolved relative phase $\varphi(x)$ between the condensates. Fourier transforming these phase profiles provides the momentum-resolved correlators ${\cal D}^{\varphi}_{k}$, which are averaged over multiple runs. These measured quantities correspond one-to-one with the theoretical spectra and 2D maps presented in the main text.

Experimentally, typical system sizes (about 100 micrometers) and imaging resolutions (a few micrometers) give access to the required momentum range. The characteristic timescales of the protocol, determined by the tunneling rate $J$ (a few Hz to tens of Hz), remain well within condensate coherence times. Shot-to-shot fluctuations in atom number translate into variations of ${\cal D}^{\varphi}_{k}$, naturally reproducing the type of inhomogeneous broadening discussed in Fig. 4 of the main text. We estimate that a few hundred to a thousand realizations are sufficient to reconstruct the 2D maps with adequate signal-to-noise, in line with current practice in phase-correlation measurements.

\section{Equations of motion}
\label{sec:eom}

Here, we start from the Hamiltonian
\begin{equation}
\label{eqn:sG_Hamiltonian}
H = \frac{1}{2} \int dx \left\lbrace \left( \partial_x \varphi \right)^2 + \Pi^2 - \left[ \Delta_0 + \delta\tilde{\Delta}_{0} (t) \right] \cos (\beta \varphi) \right\rbrace ,
\end{equation}
\sloppy where ${ \varphi }$ and ${ \Pi }$ are fields satisfying the commutation relation $\comm{\varphi(x)}{\Pi(x')}=i\delta(x-x')$, $\Delta_0$ and $\beta$ are parameters, and $ \delta\tilde{\Delta}_{0} (t) $ is our driving term. Since the latter quantity is even in the fields, only their even powers can acquire expectation value. Within a Gaussian Ansatz for the dynamics of the system~\cite{SHI_Gaussian}, we focus our attention on the two-point correlators defined as
\begin{gather}
    {\cal D}_{k}^{\varphi} (t) = \langle \delta\varphi(-k,t) \delta\varphi(k,t) \rangle, \label{eqn:D_v0_phi} \\ 
    {\cal D}_{k}^{\Pi} (t) = \langle \delta\Pi(-k,t) \delta\Pi(k,t) \rangle, \label{eqn:D_v0_Pi} \\
    {\cal D}_{k}^{X} (t) = \frac{\expval{\acomm{\delta\Pi(-k,t)}{\delta\varphi(k,t)}} + \expval{\acomm{\delta\varphi(-k,t)}{\delta\Pi(k,t)}}}{2} ,
    \label{eqn:D_v0_X}
\end{gather}
where ${\left\lbrace A, B \right\rbrace}$ indicates the anti-commutator between two operators $A$ and $B$. We note that Eqs.~\eqref{eqn:D_v0_phi}-\eqref{eqn:D_v0_X} cover the entire set of nonzero correlators in Fourier space ($x \to k$) because our driving term is spatially homogeneous. The equations of motion for these two-point observables are then given by~\cite{SHI_Gaussian}
\begin{gather}
    \partial_t {\cal D}^{\varphi}_{k}  = {\cal D}^{X}_{k} \label{eqn:dyn_2p_phi},\\
    \partial_t {\cal D}^{X}_{k}  = 2{\cal D}^{\Pi}_{k} - 2 \left\lbrace k^2 + \frac{\beta^2}{2}\left[\Delta_0+\delta\tilde{\Delta}_0(t)\right]e^{-\frac{\beta^2}{2L}\sum_q\mathcal{D}_q^\varphi}\right\rbrace \mathcal{D}^\varphi_k \label{eqn:dyn_2p_X} ,\\
    \partial_t {\cal D}^{\Pi}_{k} = -\left\lbrace k^2 + \frac{\beta^2}{2} \left[\Delta_0+\delta\tilde{\Delta}_0(t)\right]e^{-\frac{\beta^2}{2L}\sum_q\mathcal{D}_q^\varphi}\right\rbrace\mathcal{D}^X_k. \label{eqn:dyn_2p_Pi}
\end{gather}
We now proceed to write an expansion of the previous equations~\eqref{eqn:dyn_2p_phi}-\eqref{eqn:dyn_2p_Pi} in powers of the perturbation $\delta\tilde{\Delta}(t)$. The first order set of equations is given by
\begin{gather}
    \partial_t {\cal D}^{\varphi}_{k,1} - {\cal D}^{X}_{k,1} =0\label{eqn:dyn_2p_phi_1},\\
    \partial_t {\cal D}^{X}_{k,1} - 2{\cal D}^{\Pi}_{k,1} + 2 \left(k^2 + \frac{\beta^2}{2} \Delta_{\rm eq}\right)\mathcal{D}^\varphi_{k,1} - \frac{\beta^4 \Delta_{\rm eq}}{2 L} \mathcal{D}^\varphi_{k,0} \sum_q\mathcal{D}^\varphi_{q,1} = - \beta^2 \,  \delta\tilde{\Delta}(t) \, \mathcal{D}^\varphi_{k,0}  \label{eqn:dyn_2p_X_1} ,\\
    \partial_t {\mathcal D}^{\Pi}_{k,1} + \left( k^2 + \frac{\beta^2}{2} \Delta_{\rm eq}\right)\mathcal{D}^X_{k,1}=0. \label{eqn:dyn_2p_Pi_1}
\end{gather}
The sub-index $n$ in $\mathcal{D}^{\{\varphi,X,\Pi\}}_{k,n}$ indicates the order in the perturbation, with $n=0$ being the equilibrium value. Note that $\mathcal{D}^X_{k,0}=0$ due to the structure of the Hamiltonian, see Eq.~\eqref{eqn:sG_Hamiltonian}. Furthermore, we have defined $\Delta_{\rm eq}=\Delta_0\exp{-\frac{\beta^2}{2L}\sum_q \mathcal{D}^\varphi_{q,0}}$ and $\delta\tilde{\Delta}(t)=\delta\tilde{\Delta}_0(t)\exp{-\frac{\beta^2}{2L}\sum_q \mathcal{D}^\varphi_{q,0}}$. In essence, $\Delta_0$ and $\delta\tilde{\Delta}_0(t)$ are the bare coupling constants which have to be most likely regularized by the zero temperature quantum correlations of the field theory. After recasting Eqs.~\eqref{eqn:dyn_2p_phi_1}-\eqref{eqn:dyn_2p_Pi_1} in the frequency domain ($t \to \omega$), an analytical integration can be performed by self-consistently evaluating the equilibrium correlators $\mathcal{D}^\varphi_{k,0}$ and then applying a resummation scheme, which relies on the structure of the couplings between different $k$ components, to compute the first order quantity $\sum_q\mathcal{D}^\varphi_{q,1}$. The two steps are explicitly illustrated in the following sections, and the result reads
\begin{align}
    \mathcal{D}^\varphi_{k,1} (\omega) = \frac{\beta^2 \mathcal{D}_{k,0}^\varphi}{(\omega+i\eta)^2-4\left( k^2 + \frac{\beta^2}{2} \Delta_{\rm eq} \right)}\frac{\delta\tilde{\Delta}(\omega)}{1+I_1(\omega)}=\frac{1}{\eta-i\omega}\mathcal{D}^X_{k,1} (\omega),
    \label{eqn:first_order}
\end{align}
where $\eta=0^+$ is a damping coefficient and $I_1(\omega)$ is an integral which is exactly evaluated below in the $T=0$ limit. As a simple consistency check, it can be readily shown that, upon summing over all $k$ components, Eq. \eqref{eqn:first_order} yields the same expression for $\sum_k\mathcal{D}^\varphi_{k,1}$ which allowed for the integration of Eqs. \eqref{eqn:dyn_2p_phi_1}-\eqref{eqn:dyn_2p_Pi_1}.

The second order equations are given by
\begin{gather}
    \partial_t {\cal D}^{\varphi}_{k,2} - {\cal D}^{X}_{k,2} =0\label{eqn:dyn_2p_phi_2},\\
    \partial_t {\cal D}^{X}_{k,2} - 2{\cal D}^{\Pi}_{k,2} + 2 \left( k^2 + \frac{\beta^2}{2} \Delta_{\rm eq} \right) \mathcal{D}^\varphi_{k,2} - \frac{\beta^4 \Delta_{\rm eq}}{2 L} \mathcal{D}^\varphi_{k,1} \sum_q\mathcal{D}^\varphi_{q,1} + \frac{\beta^6 \Delta_{\rm eq}}{8 L^2}\mathcal{D}^\varphi_{k,0}\left(\sum_q \mathcal{D}^\varphi_{q,1} \right)^2 - \frac{\beta^4 \Delta_{\rm eq}}{2 L} \mathcal{D}^\varphi_{k,0}\sum_q \mathcal{D}^\varphi_{q,2} \notag \\ = - \beta^2 \delta\tilde{\Delta}(t) \mathcal{D}^\varphi_{k,1} + \frac{\beta^4}{2 L} \delta\tilde{\Delta}(t) \mathcal{D}^\varphi_{k,0} \sum_q \mathcal{D}^\varphi_{q,1}  \label{eqn:dyn_2p_X_2} ,\\
    \partial_t {\mathcal D}^{\Pi}_{k,2} + \left( k^2 + \frac{\beta^2}{2} \Delta_{\rm eq} \right) \mathcal{D}^X_{k,2} - \frac{\beta^4 \Delta_{\rm eq}}{4 L}\mathcal{D}^X_{k,1}\sum_q \mathcal{D}^\varphi_{q,1} = -\frac{\beta^2}{2}\delta\tilde{\Delta}(t) \mathcal{D}^X_{k,1}. \label{eqn:dyn_2p_Pi_2}
\end{gather}
Similarly to Eqs.~\eqref{eqn:dyn_2p_phi_1}-\eqref{eqn:dyn_2p_Pi_1}, we can recast Eqs.~\eqref{eqn:dyn_2p_phi_2}-\eqref{eqn:dyn_2p_Pi_2} in the frequency domain ($t\to\omega$), evaluate the second order quantity $\sum_k\mathcal{D}^\varphi_{k,2}$, see the below sections, and then solve our latest set of equations to obtain $\mathcal{D}^\varphi_{k,2} \; \forall k $. We get
\begin{equation}
\begin{split}
    \mathcal{D}^\varphi_{k,2} (\omega) & = \frac{\beta^2 \mathcal{D}_{k,0}^\varphi}{(\omega+i\eta)^2-4\left( k^2 + \frac{\beta^2}{2} \Delta_{\rm eq} \right)} \left\lbrace \left( \frac{\beta^2}{(\omega+i\eta)^2-4\left( k^2 + \frac{\beta^2}{2} \Delta_{\rm eq} \right)} \frac{\delta\tilde{\Delta}(\omega)}{1 + I_1} * \frac{\delta\tilde{\Delta}(\omega)}{1 + I_1} \right) + \right. \\
    & \left. -\frac{1}{\Delta_{\rm eq} (1 + I_1)} \left[ \left( \frac{I_2 \, \delta\tilde{\Delta}(\omega)}{1 + I_1} * \frac{\delta\tilde{\Delta}(\omega)}{1 + I_1} \right) + \frac{1}{2} \left( \frac{I_1 \, \delta\tilde{\Delta}(\omega)}{1 + I_1} * \frac{\left( 2 + I_1 \right) \delta\tilde{\Delta}(\omega)}{1 + I_1} \right) \right] + \right. \\
    & \left. + \frac{1}{-i\omega + \eta} \left[ \left( \frac{\beta^2 \left(-i\omega + \eta \right)}{(\omega+i\eta)^2-4\left( k^2 + \frac{\beta^2}{2} \Delta_{\rm eq} \right)} \frac{\delta\tilde{\Delta}(\omega)}{1 + I_1} * \frac{\delta\tilde{\Delta}(\omega)}{1 + I_1} \right) + \right. \right. \\ & \left. \left. -\frac{1}{\Delta_{\rm eq} (1 + I_1)} \left( \frac{\left(-i\omega + \eta\right)I_2 \, \delta\tilde{\Delta}(\omega)}{1 + I_1} * \frac{\delta\tilde{\Delta}(\omega)}{1 + I_1} \right) \right] \right\rbrace ,
    \label{eqn:second_order}
\end{split}
\end{equation}
where $(A*B)$ indicates the convolution operation between two functions $A$ and $B$ and $I_2(\omega, \omega')$ is an integral which is exactly evaluated below in the $T=0$ limit. Once again, self-consistency is directly confirmed by showing that, upon summation over all $k$ components, Eq.~\eqref{eqn:second_order} yields the same expression for $\sum_k\mathcal{D}^\varphi_{k,2}$ employed for integrating Eqs. \eqref{eqn:dyn_2p_phi_2}-\eqref{eqn:dyn_2p_Pi_2}. The result displayed in Eq. \eqref{eqn:second_order} can be compactified by transforming back to the time domain:
\begin{align}
    \mathcal{D}^\varphi_{k,2} (t) = \int \frac{d \omega'}{2 \pi} \int \frac{d \omega''}{2 \pi} e^{ - i \left( \omega' + \omega'' \right) t} \chi_k^{(2)} \! \left( \omega' + \omega'', \omega'' \right) \delta\tilde{\Delta}\left(\omega'\right) \delta\tilde{\Delta}\left(\omega''\right) ,
    \label{eqn:second_order_time_domain}
\end{align}
where the second order response function $ \chi_k^{(2)} \! \left( \omega' + \omega'', \omega'' \right) $ is defined as
\begin{equation}
\begin{split}
    \chi_k^{(2)} \! \left( \omega' + \omega'', \omega'' \right) &= \frac{\beta^2 \mathcal{D}_{k,0}^\varphi \left[ \left( 1 + I_1 \left( \omega' \right) \right) \left( 1 + I_1 \left( \omega'' \right) \right) \right]^{-1}}{\left(\omega' + \omega'' + i\eta \right)^2-4\left( k^2 + \frac{\beta^2}{2} \Delta_{\rm eq} \right)} \left[  \frac{\beta^2 Q\left( \omega'', \omega' \right)}{\left(\omega'' + i\eta \right)^2-4\left( k^2 + \frac{\beta^2}{2} \Delta_{\rm eq} \right)} + \right. \\ 
    & \left. \vphantom{\frac{\beta^2 Q\left( \omega'', \omega' \right)}{\left(\omega'' + i\eta \right)^2-4\left( k^2 + \frac{\beta^2}{2} \Delta_{\rm eq} \right)}}- \frac{I_2 \left( \omega' + \omega'', \omega' \right) Q\left( \omega', \omega'' \right) + \left( 1 + I_1 \left( \omega' \right) /2 \right) I_1 \left( \omega'' \right)}{\Delta_{\rm eq} \left( 1 + I_1 \left( \omega' + \omega'' \right)\right)} \right] ,
    \label{eqn: k-resolved_second_order}
\end{split}
\end{equation}
in which $ Q\left( \omega_A, \omega_B \right) = 1 + \left( - i \omega_A + \eta \right) / \left[ -i \left( \omega_A + \omega_B \right) + \eta \right] $.

\section{Equilibrium correlators}
\label{sec:eq}

We approximate the equilibrium values of the two-point correlators by employing the self-consistent mean-field approximation, which results in the following self-consistency equation
\begin{equation}
    \Omega = \frac{1}{2L}\sum_k \frac{\coth{\frac{\sqrt{k^2 + \left( \beta^2 / 2 \right) \Delta_0 \exp{-\left( \beta^2 / 2 \right) \Omega }}}{2T}}}{\sqrt{k^2 + \left( \beta^2 / 2 \right) \Delta_0 \exp{-\left( \beta^2 / 2 \right) \Omega }}} ,
    \label{eq:self-cons}
\end{equation}
where $\Omega(T) = \frac{1}{L} \sum_{k} {\cal D}_{k,0}^{\varphi} (T)$ and $L$ is the length of the system. Equation~\eqref{eq:self-cons} is not well-behaved, as usual for low-energy quantum theories, since its RHS diverges logarithmically upon integration with respect to the momentum $k$. We can regularize our theory by choosing the appropriate $\Delta_0$ resulting in the physical coupling, which we identify with $\Delta_0 e^{-\beta^2\Omega(0)/2} \equiv \Delta$. The temperature dependence is then determined by computing, in a self-consistent manner, $\delta  \Omega(T) \equiv \Omega(T)-\Omega(0)$, yielding a new (regularized) self-consistent equation
\begin{equation}
    \delta\Omega = \frac{1}{2L}\sum_k \left[ \frac{\coth{\frac{\sqrt{k^2 + \left( \beta^2 / 2 \right) \Delta \exp{-\left( \beta^2 / 2 \right) \delta \Omega }}}{2T}}}{\sqrt{k^2 + \left( \beta^2 / 2 \right) \Delta \exp{-\left( \beta^2 / 2 \right) \delta \Omega }}} - \frac{1}{\sqrt{k^2 + \left( \beta^2 / 2 \right) \Delta}} \right] .
\end{equation}
After turning the sum into an integral, the latter equation takes the form
\begin{equation}
    \delta \Omega= \frac{1}{2\pi} \int_0^\infty \left\lbrace \frac{\coth{\left[\frac{\sqrt{\left( \beta^2 / 2 \right) \Delta}}{2T}\sqrt{x^2+\exp{-\left( \beta^2 / 2 \right) \delta \Omega }}\right]}}{\sqrt{x^2+\exp{-\left( \beta^2 / 2 \right) \delta \Omega }}} - \frac{1}{\sqrt{x^2+1}}\right\rbrace dx.
    \label{eqn:domega}
\end{equation}
Upon solving this self-consistent equation for $\delta\Omega(T)$, we can compute the finite-$T$ equilibrium correlators by means of
\begin{equation}
    \mathcal{D}_{k,0}^\varphi = \frac{1}{2}\frac{\coth{\frac{\sqrt{k^2 + \left( \beta^2 / 2 \right) \Delta \exp{-\left( \beta^2 / 2 \right) \delta \Omega}}}{2T}}}{\sqrt{k^2 + \left( \beta^2 / 2 \right) \Delta \exp{-\left( \beta^2 / 2 \right) \delta \Omega}}}.
    \label{eqn: equlibrium_correlators}
\end{equation}
Finally, to ease notation we define the temperature-renormalized value of $\Delta$ as
\begin{equation}
    \Delta_{\rm eq}(T) = \Delta e^{-\frac{\beta^2}{2} \delta\Omega} .
    \label{eqn:Delta_eq}
\end{equation}

\section{Dynamics of correlators}
\label{sec:dyn}

The set of coupled linear equations~\eqref{eqn:dyn_2p_phi_1}-\eqref{eqn:dyn_2p_Pi_1} defines the propagator of the 2-point correlators around equilibrium. Said propagator can be split into a diagonal part $G_0^{-1}$, which does not mix different momenta, and an off-diagonal part $\mathcal{V}$, which does. We henceforth choose the basis defined by
\begin{align}
    \bm {\mathcal{D}}_{q,1} &= \begin{pmatrix} \mathcal{D}_{q,1}^{\varphi},\mathcal{D}_{q,1}^{X},\mathcal{D}_{q,1}^{\Pi} \end{pmatrix}^\top.
    \label{eqn: d-correl basis}
\end{align}
The diagonal part of the fluctuations' inverse propagator reads
\begin{gather}
        G_0^{-1}(\omega;q,q') = \begin{pmatrix} -i\omega + \eta & -1 & 0 \\  2 q^2 + \beta^2\Delta_{\rm eq}  & -i\omega +\eta & -2\\ 0 & q^2 + \frac{\beta^2}{2} \Delta_{\rm eq}  & -i\omega + \eta \end{pmatrix}\delta_{q,q'},
\end{gather}
where $\eta=0^+$, and the off-diagonal part is
\begin{align}
    \mathcal{V}(q,q') &= -\frac{\beta^4 \Delta_{\rm eq}}{2 L} \mathcal{D}_{q,0}^\varphi\begin{pmatrix} 0 & 0 & 0 \\
     1&0 & 0\\ 0 & 0&0 \end{pmatrix}.\label{eqn: off-diag}
\end{align}
We also define a driving vector $\bm f$, which acts as the inhomogeneous term for the set of linear equations. The driving's $q$-th component is given by
\begin{equation}
    \boldsymbol {f}_q(\omega) = - \beta^2 \, \mathcal{D}_{q,0}^{\varphi} \, \delta\tilde{\Delta}(\omega)\begin{pmatrix} 0 \\ 1 \\ 0 \end{pmatrix} .
\end{equation}
The problem at hand is to invert the matrix $G_0^{-1}+\mathcal{V}$, with components $G_0^{-1}(k,k')$ and $\mathcal{V}(k,k')$, and apply it onto the driving vector $\boldsymbol f$. Following a resummation scheme analogous to that employed in Ref.~\cite{salvador2024principles}, we arrive at
\begin{equation}
\label{eqn:chi_1}
    \frac{\beta^2 \Delta_{\rm eq}}{2L}\sum_q \mathcal{D}_{q,1}^\varphi=\frac{I_1(\omega)}{1+I_1(\omega)}\delta\tilde{\Delta}(\omega)\equiv\chi^{(1)}(\omega)\delta\tilde{\Delta}(\omega),
\end{equation}
where $\chi^{(1)} (\omega)$ is the first order response function and
\begin{equation}
    I_1(\omega)=\frac{\beta^4 \Delta_{\rm eq}}{2L}\sum_q\frac{\mathcal{D}_{q,0}^\varphi}{(\omega+i\eta)^2-4\left( q^2 + \frac{\beta^2}{2} \Delta_{\rm eq} \right)} = \frac{\beta^2}{8\pi}\int_0^\infty \frac{\coth\left[\frac{\sqrt{\left( \beta^2 / 2 \right)\Delta_{\rm eq}}}{2T}\sqrt{1+q^2}\right]}{\sqrt{1+q^2}\left[ \frac{(\omega+i\eta)^2}{2 \beta^2 \Delta_{\rm eq}}-(1+q^2) \right]} dq.
    \label{eqn:I_1}
\end{equation}
The $T=0$ limit of the $I_1$ integral is
\begin{equation}
    I_1(\omega;T=0) = \frac{\beta^2}{8\pi}\int_0^\infty \frac{1}{\sqrt{1+q^2}\left[ \frac{(\omega+i\eta)^2}{2 \beta^2 \Delta_{\rm eq}}-(1+q^2) \right]} dq ,
\end{equation}
which can be evaluated exactly for any value of $\tilde{\omega}\equiv \omega/\sqrt{2 \beta^2 \Delta_{\rm eq}}$, resulting in
\begin{equation}
I_1(\tilde{\omega};T=0) = \frac{\beta^2}{8\pi}\frac{\sqrt{-\frac{(\eta-i\tilde{\omega})^2}{1+(\eta-i\tilde{\omega})^2}}{\arcsec(\sqrt{\frac{1}{1+(\eta-i\tilde{\omega})^2}})}}{(\eta-i\tilde{\omega})^2} .
\end{equation}

The structural analogy between Eqs.~\eqref{eqn:dyn_2p_phi_1}-\eqref{eqn:dyn_2p_Pi_1} and Eqs.~\eqref{eqn:dyn_2p_phi_2}-\eqref{eqn:dyn_2p_Pi_2} allows one to exploit the above calculation approach in the second order problem as well. In the latter case, Eqs.~\eqref{eqn: d-correl basis}-\eqref{eqn: off-diag} also apply, and the driving vector is now defined as
\begin{equation}
\begin{split}
    \boldsymbol {h}_q(\omega) & = \left[ \frac{\beta^4 \Delta_{\rm eq}}{2 L} \left( \mathcal{D}_{q,1}^{\varphi} * \sum_k \mathcal{D}_{k,1}^{\varphi} \right) - \frac{\beta^6 \Delta_{\rm eq}}{8 L^2} \mathcal{D}_{q,0}^{\varphi} \left( \sum_k \mathcal{D}_{k,1}^{\varphi} * \sum_k \mathcal{D}_{k,1}^{\varphi} \right) -\beta^2 \left( \delta\tilde{\Delta}(\omega) * \mathcal{D}_{q,1}^{\varphi} \right) + \right. \\ 
    & \left. + \frac{\beta^4}{2 L} \mathcal{D}_{q,0}^{\varphi} \left( \delta\tilde{\Delta}(\omega) * \sum_k \mathcal{D}_{k,1}^{\varphi} \right) \right] \begin{pmatrix} 0 \\ 1 \\ 0 \end{pmatrix} + \left[ \frac{\beta^4 \Delta_{\rm eq}}{4 L} \left( \mathcal{D}_{q,1}^{X} * \sum_k \mathcal{D}_{k,1}^{\varphi} \right) - \frac{\beta^2}{2} \left( \delta\tilde{\Delta}(\omega) * \mathcal{D}_{q,1}^{X} \right) \right] \begin{pmatrix} 0 \\ 0 \\ 1 \end{pmatrix} ,
\end{split}
\end{equation}
where $(A*B)$ denotes the convolution operation between two functions $A$ and $B$. The resummation scheme then results in
\begin{equation}
\begin{split}
    \frac{\beta^2 \Delta_{\rm eq}}{2 L}\sum_q \mathcal{D}_{q,2}^\varphi & = \frac{1}{\Delta_{\rm eq}(1+I_1(\omega))} \left[ \left( \frac{I_2}{1 + I_1} \delta\tilde{\Delta}(\omega) * \frac{1}{1 + I_1} \delta\tilde{\Delta}(\omega) \right) + \right. \\ & \left. - \frac{I_1}{2} \left( \frac{I_1}{1 + I_1} \delta\tilde{\Delta}(\omega) * \frac{2 + I_1}{1 + I_1} \delta\tilde{\Delta}(\omega) \right) + \frac{1}{-i\omega + \eta} \left( \frac{(-i\omega + \eta) I_2}{1 + I_1} \delta\tilde{\Delta}(\omega) * \frac{1}{1 + I_1} \delta\tilde{\Delta}(\omega) \right) \right] ,
\end{split}
\end{equation}
which can be re-written as
\begin{equation}
\begin{split}
    \frac{\beta^2 \Delta_{\rm eq}}{2 L}\sum_q \mathcal{D}_{q,2}^\varphi(\omega) & = \frac{1}{\Delta_{\rm eq}(1+I_1(\omega))}\int \frac{d\omega_1}{2\pi}\int\frac{d\omega_2}{2\pi}  \left[ \frac{I_2(\omega_1,\omega)}{1 + I_1(\omega_1)} \frac{1}{1 + I_1(\omega_2)} - \frac{I_1(\omega)}{2} \frac{2 + I_1(\omega_1)}{1 + I_1(\omega_1)}  \frac{I_1(\omega_2)}{1 + I_1(\omega_2)}  \ + \right. \\ & \left.  + \frac{\omega_1+i\eta}{\omega+i\eta} \frac{I_2(\omega_1,\omega)}{1 + I_1(\omega_1)} \frac{1}{1 + I_1(\omega_2)}\right]\delta\tilde{\Delta}(\omega_1)\delta\tilde{\Delta}(\omega_2)\times2\pi\delta(\omega-\omega_1-\omega_2) .
\end{split}
\label{eqn: resummed_second_order}
\end{equation}
In the latest expressions, we have defined the quantity
\begin{equation}
\begin{split}
    I_2(\omega, \omega') & = \frac{\left( \beta^3 \Delta_{\rm eq} \right)^2}{2 L}\sum_q\frac{\mathcal{D}_{q,0}^\varphi}{\left[ (\omega+i\eta)^2-4\left( q^2 + \frac{\beta^2}{2} \Delta_{\rm eq} \right) \right] \left[ (\omega' +i\eta)^2-4\left( q^2 + \frac{\beta^2}{2} \Delta_{\rm eq} \right) \right]} \\
    & = \frac{\beta^2}{16\pi}\int_0^\infty \frac{\coth\left[\frac{\sqrt{\left( \beta^2 / 2 \right)\Delta_{\rm eq}}}{2T}\sqrt{1+q^2}\right]}{\sqrt{1+q^2}\left[ \frac{(\omega+i\eta)^2}{2 \beta^2 \Delta_{\rm eq}}-(1+q^2) \right] \left[ \frac{(\omega'+i\eta)^2}{2 \beta^2 \Delta_{\rm eq}}-(1+q^2) \right]} dq ,
\end{split}
\end{equation}
whose $T=0$ limit is given by
\begin{equation}
    I_2(\omega, \omega';T=0) = \frac{\beta^2}{16\pi}\int_0^\infty \frac{1}{\sqrt{1+q^2}\left[ \frac{(\omega+i\eta)^2}{2 \beta^2 \Delta_{\rm eq}}-(1+q^2) \right] \left[ \frac{(\omega'+i\eta)^2}{2 \beta^2 \Delta_{\rm eq}}-(1+q^2) \right]} dq.
\end{equation}
This integral can be evaluated exactly for any value of $\tilde{\omega}\equiv \omega/\sqrt{2 \beta^2 \Delta_{\rm eq}}$ and $\tilde{\omega}'\equiv \omega'/\sqrt{2 \beta^2 \Delta_{\rm eq}}$. We obtain the following result
\begin{equation}
I_2(\tilde{\omega}, \tilde{\omega}';T=0) = \frac{\beta^2}{16\pi} \frac{\frac{2 \arccos\left(\sqrt{1+(\eta -i \tilde{\omega} )^2}\right)}{\sqrt{-\left(\left(1+(\eta -i
   \tilde{\omega} )^2\right) (\eta -i \tilde{\omega} )^2\right)}}-\frac{2 \arccos \left(\sqrt{1+(\eta -i \tilde{\omega}'
   )^2}\right)}{\sqrt{-\left(\left(1+(\eta -i \tilde{\omega}' )^2\right) (\eta -i \tilde{\omega}' )^2\right)}}}{2
   (\tilde{\omega} -\tilde{\omega}' ) (\tilde{\omega} +\tilde{\omega}' +2 i \eta )} .
\end{equation}

\section{Linear response}
\begin{figure*}
    \centering
    \includegraphics[width = \textwidth]{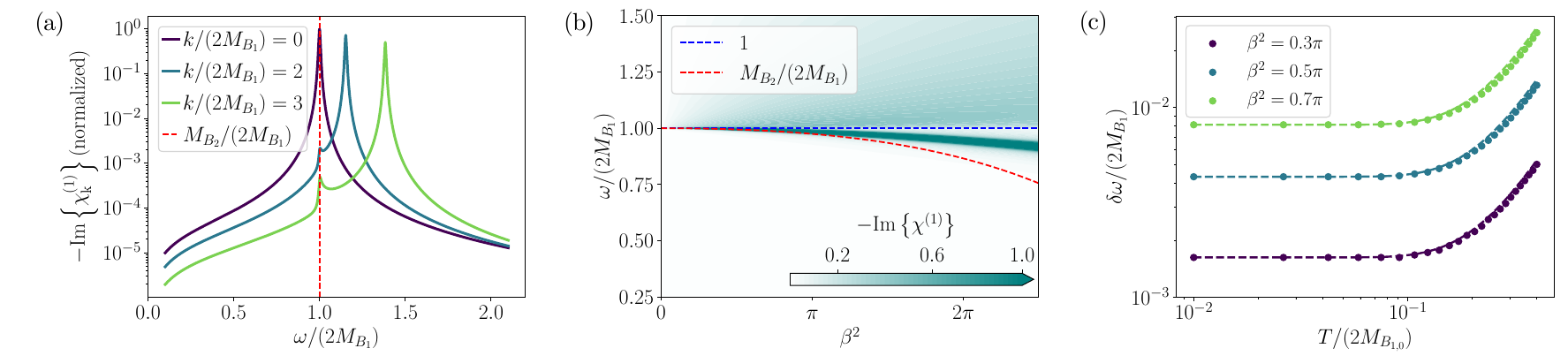}
    \caption{(a) Momentum-resolved linear response function versus frequency, for various $k$ choices. For each $k$, the left (right) peak is associated with $B_2$ ($B_1$ pair). The curves are normalized with respect to the $k = 0$ maximum, the red dashed line corresponds to the $B_2$ frequency, and we take $\beta^2 = 0.1$. (b) Linear response function, summed over $k$, versus $\beta^2$ and frequency. The blue (red) dashed line corresponds to the $2 B_1$ ($B_2$) frequency. (c) Binding energy versus temperature, for different values of $\beta^2$. The dashed lines represent the analytical result deduced from Ref.~\cite{Kazumi_Takayama_Breathers}. In panels (a) and (c), the vertical axis is logarithmic. In all panels, frequencies and energies are normalized to the $2 B_1$ mass, and a damping value $\eta = 0^+$ is used.}
    \label{fig:s3}
\end{figure*}
%
%
From the previous section, we find the linear response function $\chi_k^{(1)}(\omega)$ to be given by:
\begin{align}
    \chi_k^{(1)}(\omega) = \frac{\beta^2 \mathcal{D}_{k,0}^\varphi}{(\omega+i\eta)^2-4\left( k^2 + M_{B_1}^2 \right)}\frac{1}{1+I_1(\omega)},
    \label{eqn: first_order}
\end{align}
where we have introduced 
\begin{equation}
    I_1(\omega)=\frac{\beta^4 \Delta}{2L}\sum_q\frac{\mathcal{D}_{q,0}^\varphi}{(\omega+i\eta)^2-4\left( q^2 + M^2_{B_1} \right)}.
    \label{eqn: I_1}
\end{equation}
Here, $\mathcal{D}_{q,0}^\varphi$ are the equilibrium thermal fluctuations of field $\varphi$ and $\eta=0^+$ is a regularization that ensures the causality of the response function. In Eq.~\eqref{eqn: first_order}, we identify two contributions, see Fig.~\ref{fig:2}(b) and (c) for a schematic illustration. From left to right, the first corresponds to free $B_1$ breather pair propagation at $k$ and $-k$ momenta with a two-particle excitation gap of $2M_{B_1}$, compatible with the low-$\beta$ limit of the exact lightest breather mass. We denote this contribution by $\chi_{0k}^{(1)}(\omega)$. The second contribution, $(1+I_1)^{-1}$, corresponds to the bubble resummation of breather pairs with opposite momenta---essentially, a random phase approximation (RPA) for the response function---see the Supplemental material~\cite{Supplement} for a diagrammatic representation. This term presents a pole at $\omega_{B_2}$ fulfilling $1+{\rm{Re}}\{I_1(\omega_{B_2})\}=0$. Direct comparison with the exact solution for the breather masses~\cite{zamolodchikov1995mass} reveals the nature of this excitation as the $B_2$ breather, which is the lowest energy excitation accessible with this driving protocol~\cite{Gritsev2007}. For completeness, we report the exact expression for the masses of the $n$ breather species $B_n$, with $n = 1, \dots, [1/\xi]$:
\begin{equation}
\label{eqn:bmasses}
M_{B_n} = 2 M_S \sin \left( \frac{\pi \xi n}{2} \right) ,
\end{equation}
where $M_S$ is the soliton mass, $\xi = \beta^2 / (8\pi - \beta^2)$, and $[1/\xi]$ denotes the integer part of $1/\xi$~\cite{Gritsev2007}.



Figure~\ref{fig:s3}(a) shows the imaginary part of the momentum-resolved linear response function versus frequency for various $k$ values. With $\beta^2 = 0.1$, the $k = 0$ result exhibits a single peak near the gap frequency, characterizing both the 2-$B_1$ and $B_2$ modes in the low-$\beta$ limit (red dashed vertical line). As $k$ increases, two peaks emerge, and their separation is governed by the quadratic dispersion relation in Eq.~\eqref{eqn: first_order}. The spectral overlap between the peaks decreases with $k$, accompanied by a reduction in their intensity. For $k > 0$, the left (right) peak corresponds to the $B_2$ bound state ($B_1$ pairs with opposite momenta), with the $B_1$ pairs peak on the right being dominant.

We observe an excellent agreement between the $B_2$ breather mass computed within our Gaussian Ansatz at zero temperature and the exact solution up to values of $\beta^2 \simeq \pi$, see Fig.~\ref{fig:s3}(b). We also mention that, even if higher-energy breather excitations are expected in the full SG counterpart, our treatment captures the dominant features of the response: the 2-$B_1$ continuum and the $B_2$ breather; higher energy excitations are characterized by small oscillator strength~\cite{Gritsev2007}. Furthermore, temperature effects on the $B_2$ breather mass are inherently included in our formalism, as the $B_2$ breather emerges as the pole in the RPA resummation, which in turn depends on the equilibrium fluctuations $\mathcal{D}_{q,0}^\varphi$. We report that the binding energy of the $B_2$ breather increases with increasing phase fluctuations, and in particular with temperature, see Fig.~\ref{fig:s3}(c). We observe excellent agreement with the analytical result $ \delta \omega / ( 2 M_{B_1} ) \propto \beta^4 \coth^2{[ M_{B_1} / (2 T) ]}$ deduced from Ref.~\cite{Kazumi_Takayama_Breathers}. This is suggestive of the effective magnon-type interaction being thermally enhanced by a factor of $\coth{[ M_{B_1} / (2 T) ]}$.

\section{Two-sided Feynman diagrams for a single bosonic excitation}

The 2D protocol proposed in our work involves coupling to an operator composed of even powers of the field, i.e., $\varphi^{2n}$, rather than the usual linear coupling to $\varphi$. To get intuition on the position and intensity of the 2D spectra obtained for the SG model, here we develop a framework which generalizes the two-sided Feynman diagram rules commonly adopted in the 2D infrared spectroscopy literature~\cite{hamm_zanni_2011}. To this end, we consider the following single-boson analog of our theory ($T = 0$ here for simplicity)
\begin{equation}
    H = \frac{1}{2} \left\lbrace \Pi^2 - \left[ \Delta + \delta \tilde{\Delta} (t) \right] \cos{\left( \beta\varphi \right)} \right\rbrace .
\end{equation}
Within the Gaussian Ansatz, the effective Hamiltonian reads
\begin{equation}
    H_{G} = \frac{1}{2} \left\lbrace \Pi^2 + \left[ \Delta + \delta \tilde{\Delta} (t) \right] \frac{\beta^2}{2} e^{-\frac{\beta^2\expval{\varphi^2}}{2}} \varphi^2 \right\rbrace .
\end{equation}
To lowest order in $\beta$, introducing the relations $\varphi \sim (a + a^\dagger) / \sqrt{\Delta}$ and $\Pi \sim i(a^\dagger - a) / \sqrt{\Delta}$ with $\comm{a}{a^\dagger} = 1$, $H_{G}$ becomes the Hamiltonian of an harmonic oscillator driven by a term $ \sim \delta \tilde{\Delta} (t) (a + a^\dagger)^2 $. Omitting overall prefactors/normalizations, which are inessential in what follows, we reformulate our problem as
\begin{equation}
    H = a^\dagger a + \frac{1}{2} + D(t) (a + a^\dagger)^2 
\end{equation}
and perform a perturbative expansion of the density matrix $\rho(t)$ in powers of the driving $D(t) \ll 1$. In accordance with the analysis presented in the previous sections, we choose to focus, at the measurement step of the protocol, on the operator $(a + a^\dagger)^2 \equiv \varphi^2 $. We thus perturbatively write the system response $R(t)$ as
\begin{equation}
    R(t) = \sum_{n = 0}^\infty \Tr{\varphi^2 \rho^{(n)} (t)} \equiv \sum_{n = 0}^\infty R^{(n)} (t) ,
\end{equation}
where $\rho^{(n)} (t)$ is the $n$-th order density matrix.

The $0$-th order response, involving measurement over the unperturbed state, is readily calculated. Assuming henceforth the system to be initially prepared in the vacuum state $\ket{0}$, the density matrix is
\begin{equation}
    \rho^{(0)} (t) =
\begin{pmatrix}
1 & 0 & 0 \\
0 & 0 & 0 \\
0 & 0 & 0
\end{pmatrix} ,
\end{equation}
where we restricted our attention to the subspace spanned by the states relevant for our 2D harmonic protocol: $\left\lbrace \ket{0}, \ket{1}, \ket{2} \right\rbrace$. Since we work with $\varphi^2$ operators in our protocol, we could even safely restrict ourselves on the $2\times2$ subspace spanned by $\left\lbrace \ket{0}, \ket{2} \right\rbrace$. Given that
\begin{equation}
    \varphi^2 =
\begin{pmatrix}
1 & 0 & \sqrt{2} \\
0 & 3 & 0 \\
\sqrt{2} & 0 & 5
\end{pmatrix} ,
\end{equation}
we get
\begin{equation}
    R^{(0)} (t) = \Tr{\varphi^2 \rho^{(0)} (t)} = 1 .
\end{equation}

\begin{figure}
    \centering
    \includegraphics{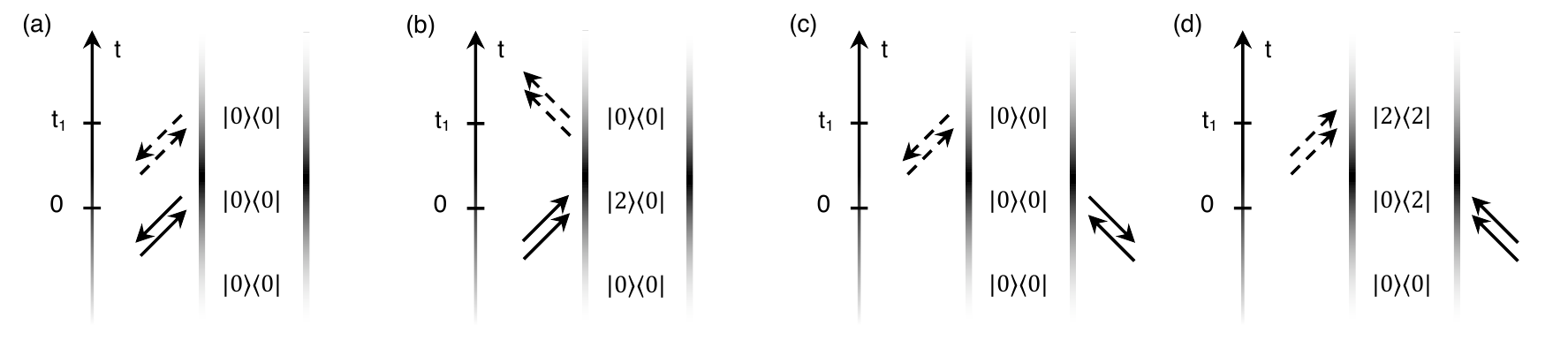}
    \caption{The four possible two-sided Feynman diagrams of linear response. Since both the interaction and the measurement involve the $\varphi^2$ operator, we draw two arrows at each step: both ingoing (outgoing) if excitation (de-excitation) takes place from a given side; mixed ingoing-outgoing if no excitation occurs from a given side. The interaction is indicated via full arrows. The measurement is always taken from the left to avoid overcounting, and it is depicted via dashed arrows. Pathways corresponding to panels (a) and (c) only differ by the side at which the interaction occurs, thus their contributions are equal and opposite and perfectly cancel. The pathways in panels (b) and (d) are responsible for the $ \chi^{(1)} (t_1) $ result given in the text.}
    \label{fig:0d-linear-diagrams}
\end{figure}

To linear order in the perturbation, we compute
\begin{equation}
    R^{(1)} (t) = \int_0^\infty d t_1 D (t - t_1) \chi^{(1)} (t_1) ,
\end{equation}
which convolves the driving pulse with the system's linear response function 
\begin{equation}
    \chi^{(1)} (t_1) = i \Tr{\varphi^2 (t_1) \comm{\varphi^2 (0)}{\rho_{\rm{eq}}}} , \label{eqn:0d-chi1}
\end{equation}
where the $\varphi^2$ operators are expressed in the Dirac picture, and $t = -\infty$ is associated with the equilibrium (initial) configuration $\rho_{\rm{eq}}$, see Refs.~\cite{Mukamel,hamm_zanni_2011} for details on the perturbative expansion of the Liouville–von Neumann equation. We now conceptually identify pathways corresponding to the terms appearing in the linear response function. We interpret the term $\varphi^2 (0)$ as the interaction with the driving pulse, occurring at time $0$, after which the system freely evolves up to time $t_1$, where a measurement of the $\varphi^2$ operator is performed. For instance, focusing on the term $ \varphi^2 (t_1) \varphi^2 (0) \rho_{\rm{eq}} $, we get
\begin{equation}
    \varphi^2 (0) \rho_{\rm{eq}} = 
    \begin{pmatrix}
1 & 0 & 0 \\
0 & 0 & 0 \\
\sqrt{2} & 0 & 0
\end{pmatrix} , \label{eqn:left-branch}
\end{equation}
that is, after the interaction the system can be thought of either remaining in a $\dyad{0}{0}$ population state or developing a $\dyad{2}{0}$ coherence. It is instructive to follow both options separately. For the $\dyad{0}{0}$ case, we have that
\begin{equation} 
    \begin{pmatrix}
1 & 0 & 0 \\
0 & 0 & 0 \\
0 & 0 & 0
\end{pmatrix} 
\xrightarrow{\text{free up to $t_1$}}
    \begin{pmatrix}
1 & 0 & 0 \\
0 & 0 & 0 \\
0 & 0 & 0
\end{pmatrix} 
\xrightarrow{\text{measure at $t_1$}}
    \begin{pmatrix}
1 & 0 & 0 \\
0 & 0 & 0 \\
\sqrt{2} & 0 & 0
\end{pmatrix} 
, \label{eqn:left-vacuum}
\end{equation}
that is, after the free vacuum evolution for time $t_1$, the measurement step brings into play a $\dyad{0}{0}$ population and a $\dyad{2}{0}$ coherence. The trace operation in Eq.~\eqref{eqn:0d-chi1} implies that contributions to our linear response arise, in the final state, from population states only. A first relevant pathway corresponds then to our system remaining in the vacuum state through both interaction and measurement. For the coherence in Eq.~\eqref{eqn:left-branch}, we instead obtain
\begin{equation} 
    \begin{pmatrix}
0 & 0 & 0 \\
0 & 0 & 0 \\
\sqrt{2} & 0 & 0
\end{pmatrix} 
\xrightarrow{\text{free up to $t_1$}}
    \begin{pmatrix}
0 & 0 & 0 \\
0 & 0 & 0 \\
e^{-i 2 t_1} \sqrt{2} & 0 & 0
\end{pmatrix} 
\xrightarrow{\text{measure at $t_1$}}
    \begin{pmatrix}
e^{-i 2 t_1} 2 & 0 & 0 \\
0 & 0 & 0 \\
e^{-i 2 t_1} 5 \sqrt{2} & 0 & 0
\end{pmatrix} 
, \label{eqn:left-coherence}
\end{equation}
that is, the second relevant pathway involves the system freely oscillating after the first interaction (up to $t_1$, thus acquiring a phase $e^{-i 2 t_1}$) and then being de-excited into the vacuum by the measurement. The term in Eq.~\eqref{eqn:0d-chi1} left to evaluate is $ \varphi^2 (t_1) \rho_{\rm{eq}} \varphi^2 (0) $. After the first interaction, now occurring “from the right”, we get
\begin{equation}
    \rho_{\rm{eq}} \varphi^2 (0) = 
    \begin{pmatrix}
1 & 0 & \sqrt{2} \\
0 & 0 & 0 \\
0 & 0 & 0
\end{pmatrix} , \label{eqn:right-branch}
\end{equation}
which can be split into
\begin{equation} 
    \begin{pmatrix}
1 & 0 & 0 \\
0 & 0 & 0 \\
0 & 0 & 0
\end{pmatrix} 
\xrightarrow{\text{free up to $t_1$}}
    \begin{pmatrix}
1 & 0 & 0 \\
0 & 0 & 0 \\
0 & 0 & 0
\end{pmatrix} 
\xrightarrow{\text{measure at $t_1$}}
    \begin{pmatrix}
1 & 0 & 0 \\
0 & 0 & 0 \\
\sqrt{2} & 0 & 0
\end{pmatrix} \label{eqn:right-vacuum}
\end{equation}
and 
\begin{equation} 
    \begin{pmatrix}
0 & 0 & \sqrt{2} \\
0 & 0 & 0 \\
0 & 0 & 0
\end{pmatrix} 
\xrightarrow{\text{free up to $t_1$}}
    \begin{pmatrix}
0 & 0 & e^{i 2 t_1} \sqrt{2} \\
0 & 0 & 0 \\
0 & 0 & 0
\end{pmatrix} 
\xrightarrow{\text{measure at $t_1$}}
    \begin{pmatrix}
0 & 0 & e^{i 2 t_1} \sqrt{2} \\
0 & 0 & 0 \\
0 & 0 & e^{i 2 t_1} 2
\end{pmatrix} 
. \label{eqn:right-coherence}
\end{equation}
The “vacuum” pathway in Eq.~\eqref{eqn:right-vacuum} gives an equal and opposite $\chi^{(1)} (t_1)$ contribution to that of Eq.~\eqref{eqn:left-vacuum}---the two interfere destructively and cancel. From Eq.~\eqref{eqn:right-coherence}, we extract a pathway where the interaction induces a $\dyad{0}{2}$ coherence, evolving freely for a time $t_1$, and the system ends up in the $\dyad{2}{2}$ population state. Overall, we get the linear response function
\begin{equation}
    \chi^{(1)} (t_1) \propto \sin \left( 2 t_1 \right) .
\end{equation}
We can visualize the pathways via two-sided Feynman diagrams, see Fig.~\ref{fig:0d-linear-diagrams}.

\begin{figure}[b!!]
    \centering
    \includegraphics{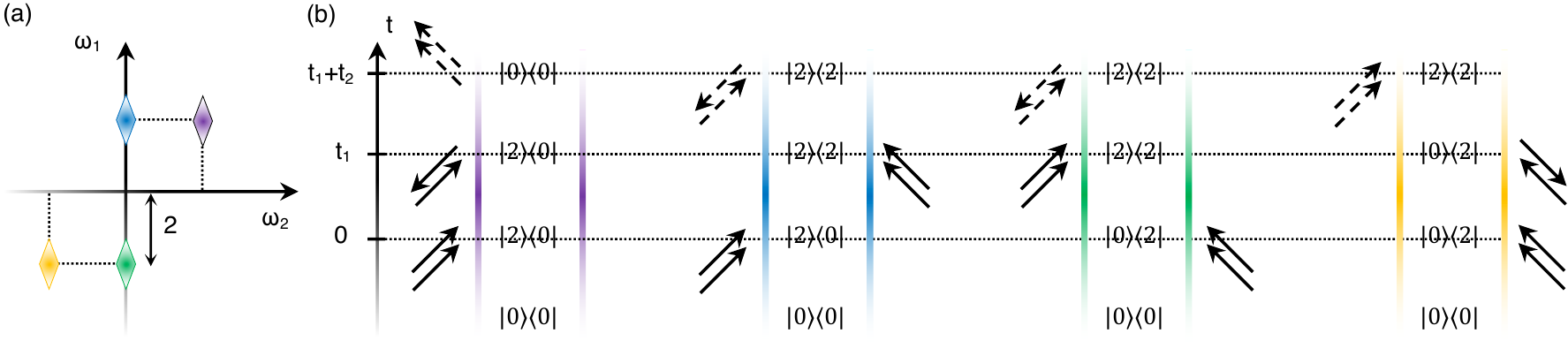}
    \caption{(a) Schematic structure of the 2D map obtained for the $\varphi^2$ protocol. (b) Two-sided Feynman diagrams corresponding to the pathways that survive destructive interference and end up dominating the 2D map. The color indicates which peak in panel~(a) a given pathway contributes to.}
    \label{fig:0d-map}
\end{figure}
We now turn to the system's $2$-nd order response, which takes the form
\begin{equation}
    R^{(2)} (t) = \int_0^\infty d t_2 \int_0^\infty d t_1 D (t - t_2) D (t - t_2 - t_1) \chi^{(2)} (t_2, t_1) ,
\end{equation}
where the system's $2$-nd order response function is
\begin{equation}
    \chi^{(2)} (t_2, t_1) = \Tr{\varphi^2 (t_1 + t_2) \comm{\varphi^2 (t_1)}{\comm{\varphi^2 (0)}{\rho_{\rm{eq}}}}} , \label{eqn:0d-chi2}
\end{equation}
which features two interactions, the first at time $0$ and the second at $t_1$, and the measurement at time $t_1 + t_2$. Through a similar analysis to that performed in linear response, we arrive at the schematic structure of Fig.~\ref{fig:0d-map}(a). Note that in this $\varphi^2$ protocol, the system always oscillates during the first time delay, i.e., no peaks are observed in the $\omega_1 = 0$ axis, and no rephasing physics occurs, i.e., the $(+,-)$ quadrant is empty. Figure~\ref{fig:0d-map}(b) schematically shows the dominant pathway (i.e., that surviving destructive interference) for each peak as a two-sided Feynman diagram. All peaks are expected with the same strength.



\section{Asymmetry of the rephasing signature}

\begin{figure}[t!!]
    \centering
    \includegraphics{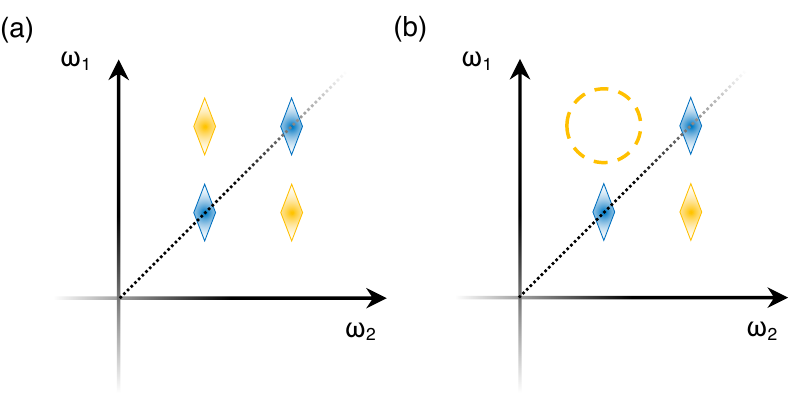}
    \caption{(a) Schematic structure of the nonrephasing spectrum for standard coupled isolated systems---the “dice-4” structure formed by two diagonal (in blue, see the dashed line) and two off-diagonal peaks (in yellow). (b) Schematic structure of the nonrephasing spectrum for a continuum of modes coupled to a single isolated mode. An off-diagonal peak is missing due to a dephasing effect on the continuum.}
    \label{fig:missing_peak}
\end{figure}

In this Appendix, we justify the absence of the upper-left off-diagonal peak in the nonrephasing sector. In standard coupled isolated systems, it is well established that the nonrephasing spectrum reveals the coupling between the modes via the appearance of two off-diagonal peaks, forming the characteristic “dice-4” pattern shown in Fig.~\ref{fig:missing_peak}(a). In contrast, our protocol involves coherent driving of a continuum of modes ($B_1$-pairs) and a single isolated mode ($B_2$ breather), which are coupled. As mentioned in the main text, the presence of the $B_1$ continuum leads to the suppression of one of the off-diagonal peaks. This results in the asymmetric signature of Fig.~\ref{fig:missing_peak}(b). We argue below that this suppression arises from the dephasing of the $B_1$ pair continuum following their excitation by the first pulse. To illustrate this, we consider the second-order response function
\begin{equation}
    \chi^{(2)}(t_1,t_2)\sim\theta(t_1)\theta(t_2)\int dx\int dy\,\Tr{\varphi_k\varphi_{-k}(t_1+t_2)\comm{\cos\varphi(t_1)}{\comm{\cos\varphi(0)}{\rho}} }.
\end{equation}
If the peak is to appear at $\omega_1~\sim E(B_{1,k}B_{1,-k})$, the first pulse drives the $B_1$-pairs, while the second pulse annihilates the continuum, and creates a $B_2$ breather (note this is a $\beta^4$ process). This means that this peak arises from
\begin{multline}
    \chi_{B_1^2,B_2}^{(2)}(t_1,t_2)\sim\theta(t_1)\theta(t_2)\int dx\int dy\,\Tr{\varphi_k\varphi_{-k}(t_1+t_2)\varphi^4(t_1)\varphi^2(0)\rho} \\ = \theta(t_1)\theta(t_2) \mel{0}{\varphi_k\varphi_{-k}}{B_2}e^{-i2\epsilon(B_{2})t_2}\times\left(\frac{1}{L}\sum_q\mel{B_2}{\varphi^4(y)}{B_{1,q}B_{1,-q}}\mel{B_{1,q}B_{1,-q}}{\varphi^2(x)}{0} e^{-i2\epsilon(B_{1,q})t_1}\right).
\end{multline}
The presence of the sum over the $q$-continuum results in a dephasing of the oscillation and the vanishing of the peak. On the other hand, the other off-diagonal peak does not vanish since
\begin{multline}
    \chi_{B_2,B_1^2}^{(2)}(t_1,t_2)\sim\theta(t_1)\theta(t_2)\Tr{\varphi_k\varphi_{-k}(t_1+t_2)\varphi^4(t_1)\varphi^2(0)\rho} \\ = \theta(t_1)\theta(t_2) \int dx\int dy\,\mel{0}{\varphi_k\varphi_{-k}}{B_{1,k}B_{1,-k}}\mel{B_{1,k}B_{1,-k}}{\varphi^4(y)}{B_2}\mel{B_2}{\varphi^2(x)}{0} e^{-i2\epsilon(B_{2})t_1} e^{-i2\epsilon(B_{1,k})t_2}
\end{multline}
does not contain the averaging over the continuum and subsequent dephasing of the oscillations.

\section{Diagrammatic understanding of the Gaussian state Ansatz}

We present in this subsection an alternative and complementary understanding of the linear response $\chi^{(1)}$ and second order response $\chi^{(2)}$ in terms of diagrammatics, following~\cite{salvador2025echo}. The action in the Keldysh contour is given by
\begin{equation}
    S[\varphi] = \int_\mathcal{C}dt\int dx\, \frac{1}{2}\left[ \left(\partial_t\varphi\right)^2-\left( \partial_x\varphi \right)^2 + \Delta_0\cos(\beta \varphi) \right],
\end{equation}
and the action corresponding to the perturbation is
\begin{align}
    S_{\rm drive}[\varphi,\delta\Delta_0]= \int_{\mathcal{C}}dt\int dx \, \frac{1}{2} \int dx \,\delta\Delta_{0} (t)  \cos (\beta \varphi).
\end{align}
Within the protocol described in the main text, the measured quantity is $\varphi^2$, which in the Keldysh path integral language corresponds to measuring $\varphi^{cl}\varphi^{cl}$. On the other hand, external perturbations correspond to the $cl-q$ representation of all the even $\varphi$ powers contained in the $\cos(\beta \varphi)$ term.

\subsection{Linear response \texorpdfstring{$\chi^{(1)}$}{x(1)}}

The linear response of the fluctutations is directly given by considering the effective $\varphi^2$ perturbation arising from the cosine perturbation and the $\varphi^2$ measurement. This directly gives to lowest order:
\begin{equation}
    \chi_0^{(1)}(\omega,k) = \vcenter{\hbox{\includegraphics[scale=0.7]{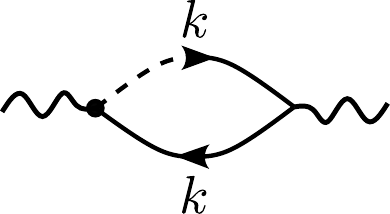}}},
\end{equation}
where the dot corresponds to the effective $\varphi^2$ arising from the cosine perturbation. The self-consistency imposes the linear response to be dressed by the resummation
\begin{equation}
    \chi^{(1)}(\omega,k) = \vcenter{\hbox{\includegraphics[scale=0.7]{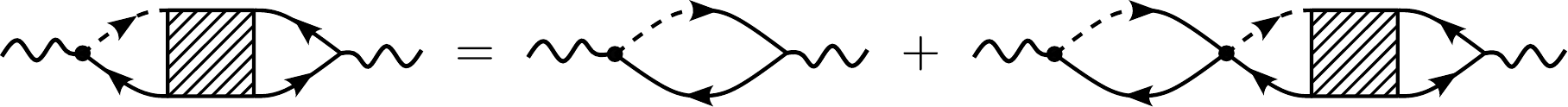}}}\,,
    \label{eqn: resumm_lin_resp}
\end{equation}
which upon solving the self-consistent equation directly results in Eq.~\eqref{eqn: first_order}. Note the resemblence with the standard density-density response for fermions, but here it is for bosons and the measurement is able to resolve the $k, -k$ pairs. For short, we define the vertex corrected bubble as 
\begin{equation}
    \chi^{(1)}(\omega,k) = \vcenter{\hbox{\includegraphics[scale=0.8]{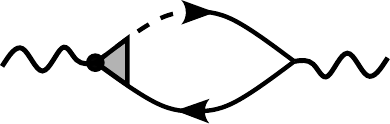}}}\,,
\end{equation}
symbolizing the resummed response given in Eq.~\eqref{eqn: resumm_lin_resp}.

\subsection{Second order response \texorpdfstring{$\chi^{(2)}$}{x(2)}}

For the nonlinear response it is instructive to first address the diagrams involved in the summed-over-$k$ response, c.f. Eq.~\eqref{eqn: resummed_second_order}. The second order is a trivial extension of the linear response, obtained by including an extra perturbation. We identify three nontrivial contributions. The first is given by
\begin{equation}
    \chi^{(2)}_1(\omega_1,\omega_2) = \vcenter{\hbox{\includegraphics[scale=0.75]{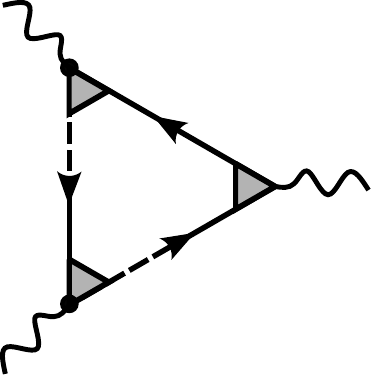}}}\sim\frac{I_2(\omega_1,\omega)}{\left[1 + I_1(\omega_1)\right]\left[1 + I_1(\omega_2)\right]\left[1 + I_1(\omega_1+\omega_2)\right]}\left(1+\frac{\omega_1+i\eta}{\omega+i\eta}\right)+\text{symmetrization}.
\end{equation}
Note that this contribution is present even in the absence of interactions, i.e., for a quadratic theory. In that case however, there would be no resummation for the 2-particle propagation---in other words, no painted triangles in the diagram. The next contribution arises from the presence of an effective $\varphi^6$ stemming from the cosine term. This diagramatically corresponds to
\begin{equation}
    \chi^{(2)}_2(\omega_1,\omega_2) = \vcenter{\hbox{\includegraphics[scale=0.75]{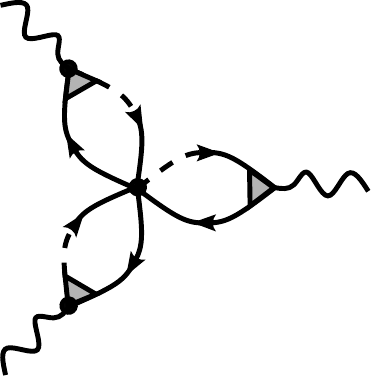}}}\sim\frac{1}{2}\frac{I_1(\omega_1)I_1(\omega_2)I_1(\omega_1+\omega_2)}{{\left[1 + I_1(\omega_1)\right]\left[1 + I_1(\omega_2)\right]\left[1 + I_1(\omega_1+\omega_2)\right]}},
\end{equation}
where the effective $\varphi^6$ interaction is given by the equilibrium fluctuations. Note this diagram would not be present if the nonlinearity would stop at $\varphi^4$. The last contribution arises from the fact that the external pertubation couples to the full cosine, not only to $\varphi^2$. This results in the possibility of processes where the pertubation is higher order in $\varphi$, and in particular within our approximation:
\begin{equation}
    \chi^{(2)}_3(\omega_1,\omega_2) = \vcenter{\hbox{\includegraphics[scale=0.75]{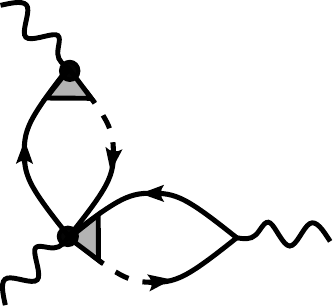}}}\sim-\frac{I_1(\omega_2)I_1(\omega_1+\omega_2)}{{\left[1 + I_1(\omega_1)\right]\left[1 + I_1(\omega_1+\omega_2)\right]}}+\text{symmetrization}.
\end{equation}
Adding the three contributions together, we readily recover Eq.~\eqref{eqn: resummed_second_order}. The $k$-resolved response can directly be obtained by imposing the two lines corresponding to the measurement (rightmost wavy line) to have $k$ momentum, recovering Eq.~\eqref{eqn: k-resolved_second_order}. The momentum-resolved response function is trivially obtained from the previous.

\end{document}